\documentclass[prd,aps,showpacs,onecolumn]{revtex4}
\usepackage{amssymb}
\usepackage{amsfonts}
\usepackage{amsmath}
\usepackage{graphicx}
\usepackage{bm}

\begin{document}

\title{Multistep DBT and regular rational extensions of the isotonic
oscillator}
\author{Yves Grandati }
\affiliation{Institut de Physique, Equipe BioPhyStat, ICPMB, IF CNRS 2843, Universit\'{e}
Paul Verlaine-Metz, 1 Bd Arago, 57078 Metz, Cedex 3, France}

\begin{abstract}
In some recent articles we developed a new systematic approach to generate
solvable rational extensions of primary translationally shape invariant
potentials. In this generalized SUSY\ QM partnership, the DBT are built on
the excited states Riccati-Schr\"{o}dinger (RS) functions regularized via
specific discrete symmetries of the considered potential. In the present
paper, we prove that this scheme can be extended in a multistep formulation.
Applying this scheme to the isotonic oscillator, we obtain new towers of
regular rational extensions of this potential which are strictly isospectral
to it. We give explicit expressions for their eigenstates which are
associated to the recently discovered exceptional Laguerre polynomials and
show explicitely that these extensions inherit of the shape invariance
properties of the original potential.
\end{abstract}

\maketitle

\section{\protect\bigskip Introduction}

Since the pionnering work of Gomez-Ullate et al \cite{gomez}, the
exceptional orthogonal polynomials (EOP) and their connexion with rational
extensions of solvable quantum potentials have been an active research
subject \cite%
{gomez2,gomez3,gomez4,gomez5,gomez6,quesne1,quesne,quesne2,quesne3,odake,sasaki,ho,odake2,sasaki2,dutta,ho2,grandati2,grandati3,grandati4}%
. The EOP appear to be the constitutive elements of the eigenstates of
solvable rational extensions of the second category primary translationally
shape-invariant potentials (TSIP) \cite{cooper,Dutt,Gendenshtein,grandati}
In a series of recent papers \cite{grandati2,grandati3,grandati4} we have
developped a new approach which allows to generate infinite sets of such
regular solvable extensions starting from every TSIP in a very direct and
systematic way without taking recourse to any ansatz or ad hoc deforming
functions. This approach is based on a generalization of the usual SUSY
partnership built from excited states. The corresponding Darboux-B\"{a}%
cklund Transformations (DBT), which are covariance transformations for the
class of Riccati-Schr\"{o}dinger (RS) equations \cite{grandati}, are based
on regularized RS functions corresponding to unphysical (negative energy)
eigenfunctions of the initial hamiltonian. They are obtained by using
discrete symmetries acting on the parameters of the considered family of
potentials. If the use of negative energy states has been already proposed
in early years of SUSY MQ development \cite{sukumar}, such a systematic
scheme has never been envisaged. In the case of the isotonic oscillator,
this gives the three infinite sets $L1$, $L2$ and $L3$ of regular rationally
solvable extensions of this potential, the $L1$ and $L2$ series potentials
being strictly isospectral to the isotonic one and inherits of its
shape-invariance property. Very recently Gomez-Ullate, Kamran and Milson 
\cite{gomez7} introduce a new family of EOP built from the classical
Laguerre polynomials via two step Darboux transformations. In the present
article we show that the scheme developed in \cite%
{grandati2,grandati3,grandati4} admits a very natural n-step extension which
allows to recover and generalize the result of \cite{gomez7}. We obtain new
infinite towers of rational extensions of the isotonic potential strictly
isospectral to this last. Their eigenfunctions are associated to the new EOP
families, the one introduced by Gomez-Ullate et al corresponding to the $n=2$
particular case.

\bigskip The paper is organized as follows. First we recall the basic
elements of the method in its one step form and the results obtained for the
isotonic system. The regularity of the RS functions is here obtained in a
self consistent way, using disconjugacy properties of the Schr\"{o}dinger
equation \cite{hartman,coppel,bocher} without recourse of the
Kienast-Lawton-Hahn theorem \cite{szego,magnus}. Then we present on a formal
level the general $m-$step scheme and in the specific case of the isotonic
oscillator, we give a criterion of concrete application in terms of
asymptotic behaviour near the origin. Next we consider in a detailed manner
the particular $m=2$ case, initially envisaged by Gomez-Ullate et al \cite%
{gomez7}. We then prove by induction a sufficient condition to be able to
build a $m-$step chain of regular and strictly isospectral extended
potentials. We finally show that the shape invariance properties of the
isotonic potential are hereditary and reached by all these extended
potentials.

During the writing of this article, two papers on the same subject appeared.
The first one \cite{odake3} contains results parallels to those developed
here. The authors adopt a point of view similar to the one initiated in \cite%
{grandati2,grandati3,grandati4} and extended in the present paper. The
so-called "virtual states deletion method" corresponding exactly to the use
of DBT based on unphysical eigenfunctions associated to regularized RS
functions. The second one \cite{quesne4} discusses quite extensively the 2
and 3 steps cases under a slightly different angle and proposes an
interesting conjecture on possible "degeneracies" in the set of the new
generated ELP. The content of the present article has been established
independently of these two works.\bigskip

\section{\protect\bigskip Darboux-B\"{a}cklund Transformations (DBT) and
discrete symmetries}

If $\psi _{\lambda }(x;a)$ is an eigenstate of $\widehat{H}%
(a)=-d^{2}/dx^{2}+V(x;a),\ a\in \mathbb{R}^{m},\ x\in I\subset \mathbb{R},$
associated to the eigenvalue $E_{\lambda }(a)$ ($E_{0}(a)=0$)

\begin{equation}
\psi _{\lambda }^{\prime \prime }(x;a)+\left( E_{\lambda }(a)-V(x;a)\right)
\psi _{\lambda }(x;a)=0,  \label{EdS}
\end{equation}
then the Riccati-Schr\"{o}dinger (RS) function $w_{\lambda }(x;a)=-\psi
_{\lambda }^{\prime }(x;a)/\psi _{\lambda }(x;a)$ satisfies the
corresponding Riccati-Schr\"{o}dinger (RS) equation \cite{grandati}

\begin{equation}
-w_{\lambda }^{\prime }(x;a)+w_{\lambda }^{2}(x;a)=V(x;a)-E_{\lambda }(a).
\label{edr4}
\end{equation}%
It is a well-known fact that the set of general Riccati equations is
invariant under the group $\mathcal{G}$ of smooth $SL(2,\mathbb{R})$-valued
curves $Map(\mathbb{R},SL(2,\mathbb{R}))$ \cite{carinena2,Ramos}. The
particular subclass of Riccati-Schr\"{o}dinger equations is, as for it,
preserved by a specific subset of $\mathcal{G}$. These transformations,
called Darboux-B\"{a}cklund Transformations (DBT), are built from any
solution $w_{\nu }(x;a)$ of the initial RS equation Eq(\ref{edr4}) as \cite%
{carinena2,Ramos,grandati}

\begin{equation}
w_{\lambda }(x;a)\overset{A\left( w_{\nu }\right) }{\rightarrow }w_{\lambda
}^{\left( \nu \right) }(x;a)=-w_{\nu }(x;a)+\frac{E_{\lambda }(a)-E_{\nu }(a)%
}{w_{\nu }(x;a)-w_{\lambda }(x;a)},  \label{transfoback2}
\end{equation}%
where $E_{\lambda }(a)>E_{\nu }(a)$. $w_{\lambda }^{\left( \nu \right) }$ is
then a solution of the RS equation:

\begin{equation}
-w_{\lambda }^{\left( \nu \right) \prime }(x;a)+\left( w_{\lambda }^{(\nu
)}(x;a)\right) ^{2}=V^{\left( \nu \right) }(x;a)-E_{\lambda }(a),
\label{eqtransform}
\end{equation}%
with the same energy $E_{\lambda }(a)$ as in Eq(\ref{edr4}) but with a
modified potential

\begin{equation}
V^{\left( \nu \right) }(x;a)=V(x;a)+2w_{\nu }^{\prime }(x;a).
\label{pottrans}
\end{equation}

This can be schematically resumed as

\begin{equation}
\left\{ 
\begin{array}{c}
w_{\lambda }\overset{A(w_{\nu })}{\rightarrowtail }w_{\lambda }^{\left( \nu
\right) } \\ 
V\overset{A(w_{\nu })}{\rightarrowtail }V^{\left( \nu \right) }.%
\end{array}%
\right.  \label{schema}
\end{equation}

The corresponding eigenstate of $\widehat{H}^{\left( \nu \right)
}(a)=-d^{2}/dx^{2}+V^{\left( \nu \right) }(x;a)$ can be written

\begin{equation}
\psi _{\lambda }^{\left( \nu \right) }(x;a)=\exp \left( -\int dxw_{\lambda
}^{(\nu )}(x;a)\right) \sim \frac{1}{\sqrt{E_{\lambda }\left( a\right)
-E_{\nu }(a)}}\widehat{A}\left( w_{\nu }\right) \psi _{\lambda }(x;a),
\label{foDBT}
\end{equation}%
where $\widehat{A}\left( a\right) $ is a first order operator given by

\begin{equation}
\widehat{A}\left( w_{\nu }\right) =d/dx+w_{\nu }(x;a).  \label{opA}
\end{equation}

Eq(\ref{foDBT}) can still be written as

\begin{equation}
\psi _{\lambda }^{\left( \nu \right) }(x;a)\sim \frac{W\left( \psi _{\nu
},\psi _{\lambda }\mid x\right) }{\psi _{\nu }(x;a)},  \label{foDBTwronsk}
\end{equation}%
where $W\left( y_{1},y_{2}\mid x\right) $ is the wronskian of the functions $%
y_{1},y_{2}$.

From $V$, the DBT generates a new potential $V^{\left( \nu \right) }$
(quasi) isospectral to the original one and its eigenfunctions are directly
obtained from those of $V$ via Eq(\ref{foDBT}). Nevertheless, in general, $%
w_{\nu }(x;a)$ and then the transformed potential $V^{\left( \nu \right)
}(x;a)$ are singular at the nodes of $\psi _{\nu }(x;a)$. For instance, if $%
\psi _{n}(x;a)$ ($\nu =n$) is a bound state of $\widehat{H}(a)$, $V^{\left(
n\right) }$ is regular only when $n=0$, that is when $\psi _{n=0}$ is the
ground state of $\widehat{H}$, and we recover the usual SUSY partnership in
quantum mechanics \cite{cooper,Dutt}.

We can however envisage to use any other regular solution of Eq(\ref{edr4})
as long as it has no zero on the considered real interval $I$, even if it
does not correspond to a physical state. For some systems, it is possible to
obtain such solutions by using specific discrete symmetries $\Gamma _{i}$
which are covariance transformations for the considered family of potentials

\begin{equation}
\left\{ 
\begin{array}{c}
a\overset{\Gamma _{i}}{\rightarrow }a_{i} \\ 
V(x;a)\overset{\Gamma _{i}}{\rightarrow }V(x;a_{i})=V(x;a)+U\left( a\right) .%
\end{array}%
\right.  \label{sym}
\end{equation}

$\Gamma _{i}$ acts on the parameters of the potential and transforms the RS
function of a\ physical excited eigenstate $w_{n}$ into a unphysical RS
function $v_{n,i}(x;a)=\Gamma _{i}\left( w_{n}(x;a)\right) =w_{n}(x;a_{i})$
associated to the negative eigenvalue $\mathcal{E}_{n,i}(a)=\Gamma
_{i}\left( E_{n}(a)\right) =U\left( a\right) -E_{n}(a_{i})<0$.

\begin{equation}
-v_{n,i}^{\prime }(x;a)+v_{n,i}^{2}(x;a)=V(x;a)-\mathcal{E}_{n,i}(a).
\label{EqRSpourv}
\end{equation}

To $v_{n,i}$ corresponds an unphysical eigenfunction of $\widehat{H}(a)$

\begin{equation}
\phi _{n,i}(x;a)=\exp \left( -\int dxv_{n,i}(x;a)\right)  \label{phi}
\end{equation}%
associated to the eigenvalue $\mathcal{E}_{n,i}(a)$.

If the transformed RS function $v_{n,i}(x;a)$ of Eq(\ref{EqRSpourv}) is
regular on $I$, it can be used to build a regular extended potential (see Eq(%
\ref{pottrans}) and Eq(\ref{foDBT}))

\begin{equation}
V^{\left( n,i\right) }(x;a)=V(x;a)+2v_{n,i}^{\prime }(x;a)
\end{equation}%
(quasi)isospectral to $V(x;a)$. The eigenstates of $V^{\left( n,i\right) }$
are given by (see Eq(\ref{transfoback2}))

\begin{equation}
\left\{ 
\begin{array}{c}
w_{k}^{\left( n,i\right) }(x;a)=-v_{n,i}(x;a)+\frac{E_{k}(a)-\mathcal{E}%
_{n,i}(a)}{v_{n,i}(x;a)-w_{k}(x;a)} \\ 
\psi _{k}^{\left( n,i\right) }(x;a)=\exp \left( -\int
dxw_{k}^{(n,i)}(x;a)\right) \sim \frac{1}{\sqrt{E_{k}\left( a\right) -%
\mathcal{E}_{n,i}(a)}}\widehat{A}\left( v_{n,i}\right) \psi _{k}(x;a),%
\end{array}%
\right.  \label{foext}
\end{equation}%
for the respective\ energies $E_{k}(a)$.\ 

The nature of the isospectrality depends if $1/\phi _{n,i}(x;a)$\ satisfies
or not the appropriate boundary conditions. If it is the case, then $1/\phi
_{n,i}(x;a)$ is a physical eigenstate of $\widehat{H}^{\left( n,i\right)
}(a)=-d^{2}/dx^{2}+V^{\left( n,i\right) }(x;a)$ for the eigenvalue $\mathcal{%
E}_{n,i}(a)$ and we only have quasi-isospectrality between $V(x;a)$\ and\ $%
V^{\left( n,i\right) }(x;a)$. If it is not the case, the isospectrality
between $V^{\left( n,i\right) }(x;a)$ and $V(x;a)$ is strict.

The above construction can be summarized by the following diagram

\begin{equation}
\begin{array}{ccccc}
E_{k}>0: & w_{k} & \overset{A(v_{n,i})}{\rightarrowtail } & w_{k}^{\left(
n,i\right) } & \left\} \text{Physical RS functions}\right. \\ 
& \Gamma _{j}\downarrow &  &  &  \\ 
\mathcal{E}_{k,j}<0: & v_{k,j} & \overset{A(v_{n,i})}{\rightarrowtail } & 
v_{k,j}^{\left( n,i\right) } & \left\} \text{ Regularized (unphysical) RS
functions}\right. \\ 
& \overset{\underbrace{}}{V} & \overset{A(v_{n,i})}{\rightarrowtail } & 
\overset{\underbrace{}}{V^{\left( n,i\right) }} & \text{Potentials}%
\end{array}
\label{diag}
\end{equation}%
with

\begin{equation}
\left\{ 
\begin{array}{c}
\psi _{k}^{\left( n,i\right) }=\widehat{A}\left( v_{n,i}\right) \psi _{k} \\ 
\phi _{k,j}^{\left( n,i\right) }=\widehat{A}\left( v_{n,i}\right) \phi
_{k,j}.%
\end{array}%
\right.  \label{DBtransfostates}
\end{equation}

This procedure can be viewed as a "generalized SUSY\ QM\ partnership" where
the DBT can be based on excited states RS\ functions properly regularized by
the symmetry $\Gamma _{j}$.

\section{Disconjugacy and regular extensions}

To control the regularity of $v_{n,i}$ we can make use of the disconjugacy
properties of the Schr\"{o}dinger equation for negative eigenvalues.

A second order differential equation like Eq(\ref{EdS}) is said to be
disconjugated on $I\subset \mathbb{R}$ ($V(x;a)$ is supposed to be
continuous on $I$) if every solution of this equation has at most one zero
on $I$ \cite{hartman,coppel}. As it is well known, this zero is necessarily
simple and at this value the considered solution changes its sign. For a
closed or open interval $I$, the disconjugacy of Eq(\ref{EdS}) is equivalent
to the existence of solutions of this equation which are everywhere non zero
on $I$ \cite{hartman,coppel}. In the following we will consider $I=\left]
0,+\infty \right[ $.

We have also the following result:

\ 

\emph{Theorem} \cite{hartman,coppel}\ If there exists a continuously
differentiable solution on $I$\ of the Riccati inequation

\begin{equation}
-w^{\prime }(x)+w^{2}(x)+G(x)\leq 0,  \label{Ricineq}
\end{equation}%
then the equation

\begin{equation}
\psi ^{\prime \prime }(x)+G(x)\psi (x)=0  \label{EdSb}
\end{equation}%
is disconjugated on $I$.

\ 

\ \ \ \ In our case, since $\mathcal{E}_{n,i}(a)\leq 0$, we have

\begin{equation}
-w_{0}^{\prime }(x;a)+w_{0}^{2}(x;a)=V(x;a)\leq V(x;a)-\mathcal{E}_{n,i}(a),
\label{inegdeconj}
\end{equation}%
$w_{0}(x;a)$ being continuously differentiable on $I$. The above theorem
ensure the existence of nodeless solutions $\phi (x;a)$ of Eq(\ref{EdS})
with $E_{\lambda }(a)=\mathcal{E}_{n,i}(a)$, that is, of regular RS
functions $v(x;a)$ solutions of Eq(\ref{EqRSpourv}). To prove that a given
solution $\phi (x;a)$ belongs to this category, it is sufficient to
determine the signs of the limit values $\phi (0^{+};a)$ and $\phi (+\infty
;a)$. If they are identical then $\phi $ is nodeless and if they are
opposite, then $\phi $ presents a unique zero on $I$.

In the first case $V(x;a)+2v^{\prime }(x;a)$ constitutes a regular
(quasi)isospectral extension of $V(x;a)$.\ \ \ \ \ \ 

In fact, B\^{o}cher \cite{bocher} has even established a more precise
theorem:

\ 

\emph{B\^{o}cher's disconjugacy theorem}

If we can find $u$\ continuously differentiable on $\left[ a,b\right] $ such
that

\begin{equation}
u^{\prime }(x)+u^{2}(x)<G(x)  \label{cond4}
\end{equation}%
on all $\left[ a,b\right] ,$\ then nor $\widetilde{\psi }(x)=u(x)\psi
(x)-\psi ^{\prime }(x)$\ nor $\psi (x)$\ can have more than one zero in $%
\left[ a,b\right] $. Moreover, if $\psi (x)$\ vanish in one point of $\left[
a,b\right] $\ then $\widetilde{\psi }(x)$\ is everywhere nonzero in this
interval (and reciprocally).

\ 

If this disconjugacy property is satisfied for every $\left[ a,b\right]
\subset \left] c,d\right[ $, then it is in fact satisfied in all the open
interval $\left] c,d\right[ $.

Consider first the case where we have $G(x)>0,\ \forall x\in \left] c,d%
\right[ $. We can then take $u(x)=0$ which implies that any solution $\psi
(x)$ of Eq(\ref{EdSb}) as well as its derivative $\psi ^{\prime }(x)$ have
at most one zero in $\left] c,d\right[ $. If $\xi _{0}$ is the corresponding
zero of $\psi ^{\prime }(x)$ then $\psi (x)$ is strictly monotonous on $%
\left] c,\xi _{0}\right[ $ and $\left] \xi _{0},d\right[ $.

If we consider now the case of Eq(\ref{EdS}) on $I=\left] 0,+\infty \right[ $
for a strictly negative eigenvalue $\mathcal{E}_{n,i}(a)<0$. From Eq(\ref%
{inegdeconj}) we can take $u(x)=-w_{0}(x;a)$ to apply B\^{o}cher's theorem
from which we deduce that not only Eq(\ref{EdS}) is disconjugate on $\left]
0,+\infty \right[ $ but also the SUSY partner $\psi ^{\left( 0\right) }(x;a)$
of any solution of Eq(\ref{EdS}) can have at most one zero on $\left]
0,+\infty \right[ $. Indeed (cf Eq(\ref{foDBTwronsk}))

\begin{equation}
\widetilde{\psi }(x;a)=\frac{W\left( \psi _{0},\psi \mid x\right) }{\psi
_{0}(x;a)}=\psi ^{\left( 0\right) }(x;a).
\end{equation}%
\ \ \ \ \ \ \ \ \ \ \ \ \ \ \ \ \ \ \ \ \ \ \ \ \ \ \ \ \ \ \ \ \ \ \ \ \ \
\ \ \ \ \ \ \ \ \ \ \ \ \ \ \ \ \ \ \ \ \ \ \ \ \ \ \ \ \ \ \ \ \ \ \ \ \ \
\ \ \ \ \ \ \ \ \ \ \ \ \ \ \ \ \ \ \ \ \ \ \ \ \ \ \ \ \ \ \ \ \ \ \ \ \ \
\ \ \ \ \ \ \ \ \ \ \ \ \ \ \ \ \ \ \ \ \ \ \ \ \ \ \ \ \ \ \ \ \ \ \ \ \ \
\ \ \ \ \ \ \ \ \ \ \ \ \ \ \ \ \ \ \ \ \ \ \ \ \ \ \ \ \ \ \ \ \ \ \ \ \ \
\ \ \ \ \ \ \ \ \ \ \ \ \ \ \ \ \ \ \ \ \ \ \ \ \ \ \ \ \ \ \ \ \ \ \ \ \ \
\ \ \ \ \ \ \ \ \ \ \ \ \ \ \ \ \ \ \ \ \ \ \ \ \ \ \ \ \ \ \ \ \ \ \ \ \ \
\ \ \ \ \ \ \ \ \ \ \ \ \ \ \ \ \ \ \ \ \ \ \ \ \ \ \ \ \ \ \ \ \ \ \ \ \ \
\ \ \ \ \ \ \ \ \ \ \ \ \ \ \ \ \ \ \ \ \ \ \ \ \ \ \ \ \ \ \ \ \ \ \ \ \ \
\ \ \ \ \ \ \ \ \ \ \ \ \ \ \ \ \ \ \ \ \ \ \ \ \ \ \ \ \ \ \ \ \ \ \ \ \ \
\ \ \ \ \ \ \ \ \ \ \ \ \ \ \ \ \ \ \ \ \ \ \ \ \ \ \ \ \ \ \ \ \ \ \ \ \ 

\section{Isotonic oscillator}

Consider the isotonic oscillator potential (ie the radial effective
potential for a three dimensional isotropic harmonic oscillator with zero
ground-state energy)

\begin{equation}
V(x;\omega ,a)=\frac{\omega ^{2}}{4}x^{2}+\frac{a(a-1)}{x^{2}}+V_{0}(\omega
,a),\ x>0,
\end{equation}%
with $a=l+1\geq 1$ and $V_{0}(\omega ,a)=-\omega \left( a+\frac{1}{2}\right) 
$. It is the unique exceptional primary translationally shape invariant
potential of the second category \cite{grandati}. The corresponding Schr\"{o}%
dinger equation is the Liouville form of the Laguerre equation on the
positive half-line and its physical spectrum, associated to the asymptotic
Dirichlet boundary conditions

\begin{equation}
\psi \left( 0^{+};\omega ,a\right) =0=\psi \left( +\infty ;\omega ,a\right)
\label{DBC}
\end{equation}%
is given by ($z=\omega x^{2}/2,$ $\alpha =a-1/2$)

\begin{equation}
\ E_{n}\left( a\right) =2n\omega ,\ \psi _{n}\left( x;\omega ,a\right)
=x^{a}e^{-z/2}\mathit{L}_{n}^{\alpha }\left( z\right) .
\end{equation}

To $\psi _{n}$ corresponds the RS function \cite{grandati}

\begin{equation}
w_{n}(x;\omega ,a)=w_{0}(x;\omega ,a)+R_{n}(x;\omega ,a),
\label{RS functions Isot}
\end{equation}%
with%
\begin{equation}
w_{0}(x;\omega ,a)=\frac{\omega }{2}x-\frac{a}{x}  \label{RS functions Isot2}
\end{equation}%
and

\begin{eqnarray}
R_{n}(x;\omega ,a) &=&\frac{-2n\omega }{\omega x-\left( 2a+1\right) /x-}\Rsh
...\Rsh \frac{2\left( n-j+1\right) \omega }{\omega x-\left( 2\left(
a+j\right) -1\right) /x-}\Rsh ...\Rsh \frac{2\omega }{\omega x-\left(
2\left( a+n\right) -1\right) /x}  \notag \\
&=&\omega x\mathit{L}_{n-1}^{\alpha +1}\left( z\right) /\mathit{L}%
_{n}^{\alpha }\left( z\right) .
\end{eqnarray}

The shape invariance property satisfied by $V$ is \cite%
{cooper,Dutt,Gendenshtein,grandati}

\begin{equation}
\widetilde{V}(x;\omega ,a)=V(x;\omega ,a)+2w_{0}^{\prime }(x;\omega
,a)=V(x;\omega ,a+1)+2\omega  \label{SIP}
\end{equation}

\bigskip We have three possible discrete symmetries for $V$ (see Eq(\ref{sym}%
)) which are given by

1)

\begin{equation}
\omega \overset{\Gamma _{+}}{\rightarrow }\left( -\omega \right) ,\left\{ 
\begin{array}{c}
V(x;\omega ,a)\overset{\Gamma _{+}}{\rightarrow }V(x;\omega ,a)+\omega (2a+1)
\\ 
w_{n}(x;\omega ,a)\overset{\Gamma _{+}}{\rightarrow }v_{n,1}(x;\omega
,a)=w_{n}(x;-\omega ,a),%
\end{array}%
\right.  \label{sym1}
\end{equation}

2)

\begin{equation}
a\overset{\Gamma _{-}}{\rightarrow }1-a,\left\{ 
\begin{array}{c}
V(x;\omega ,a)\overset{\Gamma _{-}}{\rightarrow }V(x;\omega ,a)+\omega (2a-1)
\\ 
w_{n}(x;\omega ,a)\overset{\Gamma _{-}}{\rightarrow }v_{n,2}(x;\omega
,a)=w_{n}(x;\omega ,1-a),%
\end{array}%
\right.  \label{sym2}
\end{equation}

3)

\begin{equation}
(\omega ,a)\overset{\Gamma _{3}=\Gamma _{+}\circ \Gamma _{-}}{\rightarrow }%
(-\omega ,1-a)\left\{ 
\begin{array}{c}
V(x;\omega ,a)\overset{\Gamma _{3}}{\rightarrow }V(x;\omega ,a)+2\omega \\ 
w_{n}(x;\omega ,a)\overset{\Gamma _{3}}{\rightarrow }v_{n,3}(x;\omega
,a)=w_{n}(x;-\omega ,1-a).%
\end{array}%
\right.  \label{sym3}
\end{equation}

In the $\left( \omega ,\alpha \right) $ parameters plane, $\Gamma _{+}$ and $%
\Gamma _{-}$ correspond respectively to the reflections with respect to the
axes $\omega =0$ and $\alpha =0$. The RS functions $v_{n,i},\ i=+,-,3$,
satisfy the respective RS equations

\begin{equation}
-v_{n,i}^{\prime }(x;\omega ,a)+v_{n,i}^{2}(x;\omega ,a)=V(x;\omega ,a)-%
\mathcal{E}_{n,i}(\omega ,a)\ ,  \label{EdSregL1}
\end{equation}%
with%
\begin{equation}
\left\{ 
\begin{array}{c}
\mathcal{E}_{n,+}(\omega ,a)=E_{-\left( n+a+1/2\right) }(\omega )<0 \\ 
\mathcal{E}_{n,-}(\omega ,a)=E_{n+1/2-a}\left( \omega \right) \\ 
\mathcal{E}_{n,3}(\omega ,a)=E_{-\left( n+1\right) }(\omega )<0.%
\end{array}%
\right. \   \label{Energiesneg}
\end{equation}

These eigenvalues are always negative in the $i=+$ and $i=3$ cases and the
inequality $\mathcal{E}_{n,-}(\omega ,a)\leq 0$ necessitates to satisfy the
constraint $\alpha =a-1/2>n$. When $\mathcal{E}_{n,i}(\omega ,a)\leq 0$ the
Schr\"{o}dinger equation

\begin{equation}
\phi ^{\prime \prime }(x;\omega ,a)+\left( \mathcal{E}_{n,i}(\omega
,a)-V(x;\omega ,a)\right) \phi (x;\omega ,a)=0  \label{EdS3}
\end{equation}%
is disconjugated on $\left] 0,+\infty \right[ $. We are then sure that
everywhere non zero solutions of Eq(\ref{EdS3}) exist, the question being to
if $\phi _{n,i}(x;\omega ,a)=\exp \left( -\int dxv_{n,i}(x;\omega ,a)\right) 
$ is such a solution.

\section{L1 and L2 series of extensions and corresponding exceptional
Laguerre polynomials}

The $L1$ and $L2$ series of extensions is obtained using the $\Gamma _{+}$
and $\Gamma _{-}$ symmetries respectively in which case we have synthetically

\begin{equation}
\left\{ 
\begin{array}{c}
\mathcal{E}_{n,i}\left( \omega ,a\right) =-2\omega \left( a+i\left(
n+1/2\right) \right) \\ 
\phi _{n,i}(x;\omega ,a)=\phi _{0,i}(x;\omega ,a)\mathit{L}_{n}^{i\alpha
}\left( -iz\right)%
\end{array}%
\right. ,\ i=\pm ,  \label{L12states}
\end{equation}%
with

\begin{equation}
\phi _{0,i}(x;\omega ,a)\sim x^{i\alpha +1/2}\exp \left( iz/2\right) ,
\label{L12statesb}
\end{equation}%
where $\alpha $ is supposed to statisfy the constraint $\alpha >-in$. The
corresponding RS function is given by

\begin{equation}
v_{n,i}(x;\omega ,a)=v_{0,i}(x;\omega ,a)+Q_{n,i}(x;\omega ,a),
\label{unphysstateL1}
\end{equation}%
with%
\begin{equation}
v_{0,i}(x;\omega ,a)=-i\frac{\omega }{2}x-\frac{i\alpha +1/2}{x}=-\frac{1}{x}%
\left( i\left( z+\alpha \right) +1/2\right)  \label{fondRSregL1}
\end{equation}%
and

\begin{equation}
Q_{n,i}(x;\omega ,a)=-i\omega x\mathit{L}_{n-1}^{i\alpha +1}\left(
-iz\right) /\mathit{L}_{n}^{i\alpha }\left( -iz\right) .  \label{Qn}
\end{equation}

Since \cite{szego,magnus}

\begin{equation}
\left\{ 
\begin{array}{c}
\mathit{L}_{n}^{\alpha }\left( x\right) \underset{x\rightarrow 0^{+}}{%
\rightarrow }\frac{\left( \alpha +1\right) _{n}}{n!}=\binom{n+\alpha }{n} \\ 
\mathit{L}_{n}^{\alpha }\left( x\right) \underset{x\rightarrow +\infty }{%
\sim }\frac{(-1)^{n}}{n!}x^{n},%
\end{array}%
\right.  \label{asymptLag}
\end{equation}%
where $\left( X\right) _{n}=\left( X\right) ...\left( X+n-1\right) $ is the
usual Pochhammer symbol \cite{magnus}, we have for $\alpha >1/2$

\begin{equation}
\left\{ 
\begin{array}{c}
\phi _{n,+}\left( +\infty ;\omega ,a\right) =+\infty \\ 
\phi _{n,+}\left( 0^{+};\omega ,a\right) =0^{+},%
\end{array}%
\right.  \label{asymptL1}
\end{equation}%
and for $\alpha >n$

\begin{equation}
\left\{ 
\begin{array}{c}
\phi _{n,+}\left( +\infty ;\omega ,a\right) =0^{\pm } \\ 
\phi _{n,+}\left( 0^{+};\omega ,a\right) =\pm \infty ,%
\end{array}%
\right.  \label{asymptL2}
\end{equation}%
with $\pm =\left( -1\right) ^{n}$.

Because of the constraint $\alpha >-in$, Eq(\ref{EdS3}) is disconjugated
both for $i=+$ and $i=-$. Combined to Eq(\ref{asymptL1}) and Eq(\ref%
{asymptL2}), this ensures that $\phi _{n,i}$ keeps a constant strictly
positive sign on $\left] 0,+\infty \right[ $ which means that $%
v_{n,i}(x;\omega ,a)$ is regular on this interval. Note that this result
implies in particular that is $\mathit{L}_{n}^{\alpha }\left( x\right) $
nodeless on the negative half line and that $\mathit{L}_{n}^{-\alpha }\left(
x\right) $ has no zero on the positive half line when $\alpha >n$, which is
in agreement with the Kienast-Lawton-Hahn theorem \cite{szego,magnus}.

$V^{\left( n,i\right) }(x;\omega ,a)=V(x;\omega ,a)+2v_{n,i}^{\prime
}(x;\omega ,a)$ is then a regular extension of $V(x;\omega ,a)$. Since $%
1/\phi _{n,i}\left( x;\omega ,a\right) $ diverges at the origin, it cannot
be a physical eigenstate of $\widehat{H}^{\left( n,i\right) }$.
Consequently, $V^{\left( n,i\right) }$ and $V$ are strictly isospectral.

The (unnormalized) physical eigenstates of $V^{\left( n,i\right) }$, $\psi
_{k}^{\left( n,i\right) }(x;\omega ,a)=\exp \left( -\int
dxw_{k}^{(n,i)}(x;\omega ,a)\right) $, satisfy

\begin{equation}
\left\{ 
\begin{array}{c}
\widehat{H}^{\left( n,i\right) }(\omega ,a)\psi _{k}^{\left( n,i\right)
}(x;\omega ,a)=E_{k}\left( \omega \right) \psi _{k}^{\left( n,i\right)
}(x;\omega ,a) \\ 
-w_{k}^{\left( n,i\right) \prime }(x;\omega ,a)+\left( w_{k}^{\left(
n,i\right) }(x;\omega ,a)\right) ^{2}=V^{\left( n,i\right) }(x;\omega
,a)-E_{k}\left( \omega \right) ,%
\end{array}%
\right.  \label{oregRSeq}
\end{equation}%
where, using the shape invariance property of $V$ (cf Eq(\ref{SIP})), we can
write

\begin{equation}
V^{\left( n,i\right) }(x;\omega ,a)=V(x;\omega ,a_{-1})+2Q_{n,i}^{\prime
}(x;\omega ,a).  \label{oregSUSYpart}
\end{equation}

From Eq(\ref{foext}) we have also

\begin{equation}
\psi _{k}^{\left( n,i\right) }(x;\omega ,a)=\widehat{A}\left( v_{n,i}\right)
\psi _{k}(x;\omega ,a),\ k\geq 0.
\end{equation}

More precisely, using the identities \cite{szego,magnus}

\begin{equation}
\left\{ 
\begin{array}{c}
\mathit{L}_{n}^{\alpha }\left( x\right) +\mathit{L}_{n-1}^{\alpha +1}\left(
x\right) =\mathit{L}_{n}^{\alpha +1}\left( x\right) \\ 
x\mathit{L}_{n-1}^{\alpha +1}\left( x\right) -\alpha \mathit{L}%
_{n-1}^{\alpha }\left( x\right) -n\mathit{L}_{n}^{\alpha -1}\left( x\right)
=0,%
\end{array}%
\right.  \label{recLag}
\end{equation}%
we obtain

\begin{eqnarray}
\psi _{k}^{\left( n,i\right) }(x;\omega ,a) &=&\left( v_{n,i}(x;\omega
,a)-w_{k}(x;\omega ,a)\right) \psi _{k}(x;\omega ,a)  \label{physstateL1} \\
&\sim &L_{n,k,\alpha }^{i}\left( z\right) \frac{x^{a+i}\exp (-z/2)}{%
L_{n}^{i\alpha }\left( -iz\right) },  \notag
\end{eqnarray}%
where

\begin{equation}
\left\{ 
\begin{array}{c}
L_{n,k,\alpha }^{+}\left( z\right) =L_{n}^{\alpha }\left( -z\right)
L_{k}^{\alpha +1}\left( z\right) +L_{n-1}^{\alpha +1}\left( -z\right)
L_{k}^{\alpha }\left( z\right) \\ 
L_{n,k,\alpha }^{-}\left( z\right) =\left( k+n+\alpha \right) L_{k}^{\alpha
}\left( z\right) L_{n}^{-\alpha }\left( z\right) -\left( -n+\alpha \right)
L_{k}^{\alpha }\left( z\right) L_{n-1}^{-\alpha }\left( z\right) -\left(
k+\alpha \right) L_{k-1}^{\alpha }\left( z\right) L_{n}^{-\alpha }\left(
z\right) ,%
\end{array}%
\right.  \label{ELP12}
\end{equation}%
are polynomials of degree $n+k$, namely exceptional Laguerre polynomials
(ELP) of the $L1$ and $L2$ series respectively. From the orthogonality
conditions on the eigenstates for fixed values of $n$ and $a$, we retrieve
the fact that the ELP $L_{n,k,\alpha }^{i}\left( z\right) $ constitute
orthogonal families with the corresponding weights

\begin{equation}
W_{n}^{i}\left( z\right) =\frac{z^{a+i/2}\exp (-z)}{\left( L_{n}^{i\alpha
}\left( -iz\right) \right) ^{2}}.  \label{weight12}
\end{equation}

\section{Formal scheme for multi-step DBT}

\subsection{General elements}

The question of successive iterations of DBT is very natural and is at the
center of the construction of the hierarchy of hamiltonians in the usual
SUSY QM scheme \cite{sukumar2}. Staying at the formal level, it can be
simply described by the following straightforward generalization of Eq(\ref%
{schema})

\begin{equation}
\left\{ 
\begin{array}{c}
w_{\lambda }\overset{A(w_{\nu _{1}})}{\rightarrowtail }w_{\lambda }^{\left(
\nu _{1}\right) }\overset{A(w_{\nu _{2}}^{\left( \nu _{1}\right) })}{%
\rightarrowtail }w_{\lambda }^{\left( \nu _{1},\nu _{2}\right) }...\overset{%
A(w_{\nu _{m}}^{\left( \nu _{1},,...,\nu _{m-1}\right) })}{\rightarrowtail }%
w_{\lambda }^{\left( \nu _{1},...,\nu _{m}\right) } \\ 
V\overset{A(w_{\nu _{1}})}{\rightarrowtail }V^{\left( \nu _{1}\right) }%
\overset{A(w_{\nu _{2}}^{\left( \nu _{1}\right) })}{\rightarrowtail }%
V^{\left( \nu _{1},\nu _{2}\right) }...\overset{A(w_{\nu _{m}}^{\left( \nu
_{1},,...,\nu _{m-1}\right) })}{\rightarrowtail }V^{\left( \nu _{1},...,\nu
_{m}\right) },%
\end{array}%
\right.  \label{diagn}
\end{equation}%
where $w_{\lambda }^{\left( \nu _{1},...,\nu _{m}\right) }$ is a RS function
associated to the eigenvalue $E_{\lambda }$ of the potential

\begin{equation}
V^{\left( \nu _{1},...,\nu _{m}\right) }(x;a)=V(x;a)+2\sum_{j=1}^{m-1}\left(
w_{\nu _{j}}^{\left( \nu _{1},,...,\nu _{j-1}\right) }(x;a)\right) ^{\prime
}.  \label{potnstep}
\end{equation}

The corresponding eigenfunction is given by (cf Eq(\ref{foDBT}) and Eq(\ref%
{foDBTwronsk}))

\begin{equation}
\psi _{\lambda }^{\left( \nu _{1},...,\nu _{m}\right) }(x;a)=\widehat{A}%
\left( w_{\nu _{m}}^{\left( \nu _{1},...,\nu _{m-1}\right) }\right) \psi
_{\lambda }^{\left( \nu _{1},...,\nu _{m-1}\right) }(x;a)=\widehat{A}\left(
w_{\nu _{m}}^{\left( \nu _{1},...,\nu _{m-1}\right) }\right) ...\widehat{A}%
\left( w_{\nu _{1}}\right) \psi _{\lambda }(x;a),  \label{etats n}
\end{equation}%
that is,

\begin{eqnarray}
\psi _{\lambda }^{\left( \nu _{1},...,\nu _{m}\right) }(x;a) &=&\left(
w_{\nu _{m}}^{\left( \nu _{1},...,\nu _{m-1}\right) }(x;a)-w_{\lambda
}^{\left( \nu _{1},...,\nu _{m-1}\right) }(x;a)\right) \psi _{\lambda
}^{\left( \nu _{1},...,\nu _{m-1}\right) }(x;a)  \label{etats n2} \\
&=&\frac{W\left( \psi _{\nu _{m}}^{\left( \nu _{1},...,\nu _{m-1}\right)
},\psi _{\lambda }^{\left( \nu _{1},...,\nu _{m-1}\right) }\mid x\right) }{%
\psi _{\nu _{m}}^{\left( \nu _{1},...,\nu _{m-1}\right) }(x;a)},  \notag
\end{eqnarray}%
where $W\left( y_{1},...,y_{m}\mid x\right) $ is the wronskian of the
functions $y_{1},...,y_{m}$

\begin{equation}
W\left( y_{1},...,y_{m}\mid x\right) =\left\vert 
\begin{array}{ccc}
y_{1}\left( x\right) & ... & y_{m}\left( x\right) \\ 
... &  & ... \\ 
y_{1}^{\left( m-1\right) }\left( x\right) & ... & y_{m}^{\left( m-1\right)
}\left( x\right)%
\end{array}%
\right\vert .  \label{wronskien}
\end{equation}

Other useful representations for the extended potentials and of their
eigenfunctions are possible. From Sylvester's theorem we can deduce \cite%
{muir} the following property for the Wronskians

\begin{equation}
W\left( y_{1},...,y_{m},y\mid x\right) =\frac{W\left( W\left(
y_{1},...,y_{m}\right) ,W\left( y_{1},...,y_{m-1},y\right) \mid x\right) }{%
W\left( y_{1},...,y_{m-1}\mid x\right) }.  \label{sylv}
\end{equation}

Using the identity \cite{muir}

\begin{equation}
W\left( uy_{1},...,uy_{m}\mid x\right) =u^{m}W\left( y_{1},...,y_{m}\mid
x\right) ,  \label{factwronsk}
\end{equation}%
this gives

\begin{equation}
\frac{W\left( y_{1},...,y_{m},y\mid x\right) }{W\left( y_{1},...,y_{m}\mid
x\right) }=\frac{W\left( \frac{W\left( y_{1},...,y_{m}\right) }{W\left(
y_{1},...,y_{m-1}\right) },\frac{W\left( y_{1},...,y_{m-1},y\right) }{%
W\left( y_{1},...,y_{m-1}\right) }\mid x\right) }{\frac{W\left(
y_{1},...,y_{m}\mid x\right) }{W\left( y_{1},...,y_{m-1}\mid x\right) }}
\end{equation}%
and comparing to Eq(\ref{etats n2}), we obtain

\begin{equation}
\psi _{\lambda }^{\left( \nu _{1},...,\nu _{m}\right) }(x;a)=\frac{W\left(
\psi _{\nu _{1}},...,\psi _{\nu _{m}},\psi _{\lambda }\mid x\right) }{%
W\left( \psi _{\nu _{1}},...,\psi _{\nu _{m}}\mid x\right) },
\label{etats n3}
\end{equation}%
which is the well known Crum formula \cite{crum,matveev} for the
eigenfunctions.

Inserting this result in Eq(\ref{potnstep}), we then deduce the Crum formula
for the potential \cite{crum,matveev}

\begin{equation}
V^{\left( \nu _{1},...,\nu _{m}\right) }(x;a)=V(x;a)+2\left( \log W\left(
\psi _{\nu _{1}},...,\psi _{\nu _{m}}\mid x\right) \right) ^{\prime \prime }.
\label{potnstep2}
\end{equation}

Since all the functions $\psi _{\nu _{j}}$ implied in the wronskians in Eq(%
\ref{etats n3}) and Eq(\ref{potnstep2}) are eigenfunctions of the same
hamiltonian $\widehat{H}(a)$, the properties of the determinants allow us to
replace in these wronskians the even derivatives $\psi _{\nu _{j}}^{\left(
2l\right) }$ by $\left( -E_{\nu _{j}}\right) ^{l}\psi _{\nu _{j}}$ and the
odd derivatives $\psi _{\nu _{j}}^{\left( 2l+1\right) }$ by $\left( -E_{\nu
_{j}}\right) ^{l}\psi _{\nu _{j}}^{\prime }$.

For instance

\begin{eqnarray}
W\left( \psi _{\nu _{1}},...,\psi _{\nu _{2l}}\mid x\right)  &=&\left\vert 
\begin{array}{ccc}
\psi _{\nu _{1}}\left( x;a\right)  & ... & \psi _{\nu _{2l}}\left(
x;a\right)  \\ 
\psi _{\nu _{1}}^{\prime }\left( x;a\right)  & ... & \psi _{\nu
_{2l}}^{\prime }\left( x;a\right)  \\ 
... &  & ... \\ 
\left( -E_{\nu _{1}}\right) ^{l-1}\psi _{\nu _{1}}\left( x;a\right)  & ... & 
\left( -E_{\nu _{2l}}\right) ^{l-1}\psi _{\nu _{2l}}\left( x;a\right)  \\ 
\left( -E_{\nu _{1}}\right) ^{l-1}\psi _{\nu _{1}}^{\prime }\left(
x;a\right)  & ... & \left( -E_{\nu _{2l}}\right) ^{l-1}\psi _{\nu
_{2l}}^{\prime }\left( x;a\right) 
\end{array}%
\right\vert   \label{CKdeteven} \\
&=&\Delta ^{\left( \nu _{1},...,\nu _{2l}\right) }\left( x;a\right)
\prod_{j=1}^{2l}\psi _{\nu _{j}}\left( x;a\right) ,  \notag
\end{eqnarray}%
where

\begin{equation}
\Delta ^{\left( \nu _{1},...,\nu _{2l}\right) }\left( x;a\right)
=(-1)^{l}\left\vert 
\begin{array}{ccc}
1 & ... & 1 \\ 
w_{\nu _{1}}\left( x;a\right)  & ... & w_{\nu _{2l}}\left( x;a\right)  \\ 
... &  & ... \\ 
\left( E_{\nu _{1}}\right) ^{l-1} & ... & \left( E_{\nu _{2l}}\right) ^{l-1}
\\ 
\left( E_{\nu _{1}}\right) ^{l-1}w_{\nu _{1}}\left( x;a\right)  & ... & 
\left( E_{\nu _{2l}}\right) ^{l-1}w_{\nu _{2l}}\left( x;a\right) 
\end{array}%
\right\vert   \label{CKdeteven1}
\end{equation}%
and%
\begin{eqnarray}
W\left( \psi _{\nu _{1}},...,\psi _{\nu _{2l+1}}\mid x\right)  &=&\left\vert 
\begin{array}{ccc}
\psi _{\nu _{1}}\left( x;a\right)  & ... & \psi _{\nu _{2l}}\left(
x;a\right)  \\ 
\psi _{\nu _{1}}^{\prime }\left( x;a\right)  & ... & \psi _{\nu
_{2l}}^{\prime }\left( x;a\right)  \\ 
... &  & ... \\ 
\left( -E_{\nu _{1}}\right) ^{l-1}\psi _{\nu _{1}}^{\prime }\left(
x;a\right)  & ... & \left( -E_{\nu _{2l+1}}\right) ^{l-1}\psi _{\nu
_{2l+1}}^{\prime }\left( x;a\right)  \\ 
\left( -E_{\nu _{1}}\right) ^{l}\psi _{\nu _{1}}\left( x;a\right)  & ... & 
\left( -E_{\nu _{2l+1}}\right) ^{l}\psi _{\nu _{2l+1}}\left( x;a\right) 
\end{array}%
\right\vert   \label{CKdetodd} \\
&=&\Delta ^{\left( \nu _{1},...,\nu _{2l+1}\right) }\left( x;a\right)
\prod_{j=1}^{2l+1}\psi _{\nu _{j}}\left( x;a\right) ,  \notag
\end{eqnarray}%
where

\begin{equation}
\Delta ^{\left( \nu _{1},...,\nu _{2l+1}\right) }\left( x;a\right)
=\left\vert 
\begin{array}{ccc}
1 & ... & 1 \\ 
w_{\nu _{1}}\left( x;a\right) & ... & w_{\nu _{2l+1}}\left( x;a\right) \\ 
... &  & ... \\ 
\left( E_{\nu _{1}}\right) ^{l-1}w_{\nu _{1}}\left( x;a\right) & ... & 
\left( E_{\nu _{2l}}\right) ^{l-1}w_{\nu _{2l+1}}\left( x;a\right) \\ 
\left( E_{\nu _{1}}\right) ^{l} & ... & \left( E_{\nu _{2l}+1}\right) ^{l}%
\end{array}%
\right\vert .  \label{CKdetodd1}
\end{equation}

Then we obtain the representation of the $\psi _{\lambda }^{\left( \nu
_{1},...,\nu _{m}\right) }$ in terms the so-called Crum-Krein determinants $%
\Delta ^{\left( \nu _{1},...,\nu _{m}\right) }$ \cite{crum,krein,bagrov}

\begin{equation}
\psi _{\lambda }^{\left( \nu _{1},...,\nu _{m}\right) }(x;a)=\psi _{\lambda
}\left( x;a\right) \frac{\Delta ^{\left( \nu _{1},...,\nu _{m},\lambda
\right) }\left( x;a\right) }{\Delta ^{\left( \nu _{1},...,\nu _{m}\right)
}\left( x;a\right) }.  \label{CKdet}
\end{equation}

Until now, all these results are purely formal. The central problem is now
to choose in an appropriate way the family of eigenfunctions $\left( \psi
_{\nu _{1}},...,\psi _{\nu m}\right) $ from which the successive DBT are
built (see Eq(\ref{etats n3}) and Eq(\ref{potnstep})) in order to ensure the
regularity of the successive extended potentials $V^{\left( \nu _{1},...,\nu
_{m}\right) }$.

A first answer has been given more than half a century ago by Crum \cite%
{crum}. The Crum proposal \cite{crum} is to take the successive eigenstates
of the discrete bound states spectrum starting from the ground state

\begin{equation}
\left( \nu _{1},...,\nu _{m}\right) =\left( 0,...,m-1\right)
\end{equation}

Krein \cite{krein,bagrov,bagrov1} extended this result showing it is
possible to choose sets of unnecessarily successive eigenstates if they
satisfy a given condition, namely

\begin{equation}
\left( \nu _{1},...,\nu _{m}\right) =\left( n_{1},...,n_{m}\right) \in 
\mathbb{N}^{m},
\end{equation}%
with

\begin{equation}
(n-n_{1})(n-n_{2})...(n-n_{m})\geq 0,\ \forall n\in \mathbb{N}.
\end{equation}

This last is verified if the set $\left( n_{1},...,n_{m}\right) $ is
constituted by "aggregates" of an even number of eigenstates. The
possibility to employ sets of two juxtaposed eigenstates has been
rediscovered later by Adler \cite{adler}. These results have been used
extensively in the context of higher order SUSY (see for instance \cite%
{sukumar,sukumar2,andrianov,andrianov2,bagrov,bagrov1,samsonov,samsonov1,samsonov2,fernandez,fernandez2,fernandez3,mielnik}%
). In \cite{samsonov2}, Samsonov has still extended the Krein-Adler result.
Note finally that Fernandez et al \cite{fernandez} employed the specific
"Wick rotated" eigenfunctions of negative energies introduced by Shnol' \cite%
{shnol',samsonov,grandati2} to build successive extensions of the harmonic
potential.

\subsection{Application to the isotonic oscillator}

Our aim is to consider the possibility to generalize our construction in a
multi-step version on the basis of the regularized RS functions (or
unphysical eigenfunctions) obtained above. Staying at a formal level, it
would correspond to the following generalization of the diagram Eq(\ref{diag}%
) (see Eq(\ref{diagn}))

\begin{equation}
\begin{array}{ccccccccc}
E_{k}>0: & w_{k} & \overset{A(v_{n_{1},i_{1}})}{\rightarrowtail } & 
w_{k}^{\left( n_{1},i_{1}\right) } & \overset{A(v_{n_{2},i_{2}}^{\left(
n_{1},i_{1}\right) })}{\rightarrowtail } & w_{k}^{\left(
n_{1},i_{1},n_{2},i_{2}\right) } & ... & \overset{A(v_{n_{m},i_{m}}^{\left(
n_{1},i_{1},,...,n_{m-1},i_{m-1}\right) })}{\rightarrowtail } & 
w_{k}^{\left( n_{1},i_{1},,...,n_{m},i_{m}\right) } \\ 
& \Gamma _{i}\downarrow &  &  &  &  &  &  &  \\ 
\mathcal{E}_{k,i}<0: & v_{k,i} & \overset{A(v_{n_{1},i_{1}})}{%
\rightarrowtail } & v_{k,i}^{\left( n_{1},i_{1}\right) } & \overset{%
A(v_{n_{2},i_{2}}^{\left( n_{1},i_{1}\right) })}{\rightarrowtail } & 
v_{k,i}^{\left( n_{1},i_{1},n_{2},i_{2}\right) } & ... & \overset{%
A(v_{n_{m},i_{m}}^{\left( n_{1},i_{1},,...,n_{m-1},i_{m-1}\right) })}{%
\rightarrowtail } & v_{k,i}^{\left( n_{1},i_{1},,...,n_{m},i_{m}\right) } \\ 
& \overset{\underbrace{\qquad }}{V} & \overset{A(v_{n_{1},i_{1}})}{%
\rightarrowtail } & \overset{\underbrace{\qquad \qquad }}{V^{\left(
n_{1},i_{1}\right) }} & \overset{A(v_{n_{2},i_{2}}^{\left(
n_{1},i_{1}\right) })}{\rightarrowtail } & \overset{\underbrace{\qquad
\qquad \qquad }}{V^{\left( n_{1},i_{1},n_{2},i_{2}\right) }} & ... & \overset%
{A(v_{n_{m},i_{m}}^{\left( n_{1},i_{1},,...,n_{m-1},i_{m-1}\right) })}{%
\rightarrowtail } & \overset{\underbrace{\qquad \qquad \qquad \qquad }}{%
V^{\left( n_{1},i_{1},,...,n_{m},i_{m}\right) }}%
\end{array}
\label{diagniso}
\end{equation}%
with $v_{k}^{\left( n_{0}\right) }=v_{k}$. Under this way, we generate
chains of isospectral extensions $V^{\left(
n_{1},i_{1},,...,n_{m},i_{m}\right) }$ of the potential $V$ with (see Eq(\ref%
{potnstep}) and Eq(\ref{potnstep2}))

\begin{eqnarray}
V^{\left( n_{1},i_{1},...,n_{m},i_{m}\right) }(x;a) &=&V(x;a)+2\left(
v_{n_{m},i_{m}}^{\left( n_{1},i_{1},...,n_{m-1},i_{m-1}\right) }(x;a)\right)
^{\prime }  \label{potnstepisot} \\
&=&V(x;a)+2\sum_{j=1}^{m}\left( v_{n_{j},i_{j}}^{\left(
n_{1},i_{1},...,n_{j-1},i_{j-1}\right) }(x;a)\right) ^{\prime }  \notag \\
&=&V(x;a)+2\left( \log W\left( \phi _{n_{1},i_{1}},...,\phi
_{n_{m},i_{m}}\mid x\right) \right) ^{\prime \prime }  \notag
\end{eqnarray}

$w_{k}^{\left( n_{1},i_{1},,...,n_{m},i_{m}\right) }$ and the $%
v_{k,i}^{\left( n_{1},i_{1},,...,n_{m},i_{m}\right) }$ are RS functions of
the extended potential $V^{\left( n_{1},i_{1},...,n_{m},i_{m}\right) }$
associated respectively to the eigenvalues $E_{k}$ and $\mathcal{E}_{k,i}$.

The corresponding eigenfunctions are given by (see Eq(\ref{etats n2}) and Eq(%
\ref{etats n3}))

\begin{equation}
\left\{ 
\begin{array}{c}
\psi _{k}^{\left( n_{1},i_{1},...,n_{m},i_{m}\right) }(x;a)=\frac{W\left(
\phi _{n_{m},i_{m}}^{\left( n_{1},i_{1},...,n_{m-1},i_{m-1}\right) },\psi
_{k}^{\left( n_{1},i_{1},...,n_{m-1},i_{m-1}\right) }\mid x\right) }{\phi
_{n_{m},i_{m}}^{\left( n_{1},i_{1},...,n_{m-1},i_{m-1}\right) }(x;a)}=\frac{%
W\left( \phi _{n_{1},i_{1}},...,\phi _{n_{m},i_{m}},\psi _{k}\mid x\right) }{%
W\left( \phi _{n_{1},i_{1}},...,\phi _{n_{m},i_{m}}\mid x\right) } \\ 
\phi _{k,i}^{\left( n_{1},i_{1},...,n_{m},i_{m}\right) }(x;a)=\frac{W\left(
\phi _{n_{m},i_{m}}^{\left( n_{1},i_{1},...,n_{m-1},i_{m-1}\right) },\phi
_{k,i}^{\left( n_{1},i_{1},...,n_{m-1},i_{m-1}\right) }\mid x\right) }{\phi
_{n_{m},i_{m}}^{\left( n_{1},i_{1},...,n_{m-1},i_{m-1}\right) }(x;a)}=\frac{%
W\left( \phi _{n_{1},i_{1}},...,\phi _{n_{m},i_{m}},\phi _{k,i}\mid x\right) 
}{W\left( \phi _{n_{1},i_{1}},...,\phi _{n_{m},i_{m}}\mid x\right) }.%
\end{array}%
\right.  \label{etats niso}
\end{equation}

Suppose that at the step $m-1$ all the extended potentials in the chain are
regular and strictly isospectral. The problem is now to control the
regularity of the potential $V^{\left( n_{1},i_{1},...,n_{m},i_{m}\right) }$
and of its physical eigenstates $\psi _{k}^{\left(
n_{1},i_{1},,...,n_{m-1},i_{m-1}\right) }$ as well as its strict
isospectrality with $V^{\left( n_{1},i_{1},...,n_{m-1},i_{m-1}\right) }$.
This is achieved if the unphysical eigenfunction $\phi
_{n_{m},i_{m}}^{\left( n_{1},i_{1},,...,n_{m-1},i_{m-1}\right) }$,
associated to the DBT $A(v_{n_{m},i_{m}}^{\left(
n_{1},i_{1},,...,n_{m-1},i_{m-1}\right) })$, is nodeless and if $1/\phi
_{n_{m},i_{m}}^{\left( n_{1},i_{1},,...,n_{m-1},i_{m-1}\right) }$ cannot be
into the set of physical eigenstates of $V^{\left(
n_{1},i_{1},...,n_{m-1},i_{m-1}\right) }$.

The disconjugacy of the Schr\"{o}dinger equation

\begin{equation}
\phi ^{\prime \prime }(x;\omega ,a)+\left( \mathcal{E}_{n_{m},i_{m}}(\omega
,a)-V^{\left( n_{1},i_{1},...,n_{m-1},i_{m-1}\right) }(x;\omega ,a)\right)
\phi (x;\omega ,a)=0,
\end{equation}%
for a negative eigenvalue $\mathcal{E}_{n_{m},i_{m}}$ guarantees that $%
v_{n_{m},i_{m}}^{\left( n_{1},i_{1},...,n_{m-1},i_{m-1}\right) }$ is at most
singular in one point. The regularity of the extended potential $V^{\left(
n_{1},i_{1},...,n_{m},i_{m}\right) }$ and its strict isospectrality with the
preceding one in the chain are then satisfied as soon as the\ unphysical
eigenfunction $\phi _{n_{m},i_{m}}^{\left(
n_{1},i_{1},,...,n_{m-1},i_{m-1}\right) }$ satisfies appropriate boundary
conditions, which was already the argument used above in the one step case.
Namely, the strict isospectrality and the regularity are satisfied when $%
\phi _{n_{m},i_{m}}^{\left( n_{1},i_{1},...,n_{m-1},i_{m-1}\right) }$ tends
to $\infty $ at one extremity of $\left] 0,+\infty \right[ $ and to $0$ at
the other extremity, with the same sign for both limits.

In the case of the isotonic oscillator, there are in fact three type of
regularization transformations $\Gamma _{i},\ i=+,-,3$ associated
respectively to the three series of EOP $L1$, $L2$ and $L3$ \cite%
{quesne,grandati3}. At each step we can a priori use any of the three type
if regularized RS function. However, since the use of DBT associated to RS
functions of the $L3$ series leads only to quasi isospectrality, the
argument above is no more valid. Consequently, in the following we exclude
this case and consider only DBT built from the RS functions of the $L1$ or $%
L2$ series.

From Eq(\ref{CKdet}), we see that in the present case\bigskip 
\begin{equation}
\phi _{n_{m},i_{m}}^{\left( n_{1},i_{1},...,n_{m-1},i_{m-1}\right)
}(x;\omega ,a)=\phi _{0,i_{m}}\left( x;\omega ,a\right) R^{\left(
n_{1},i_{1},,...,n_{m},i_{m}\right) }\left( x;\omega ,a\right) ,
\end{equation}%
where

\begin{equation}
R^{\left( n_{1},i_{1},,...,n_{m},i_{m}\right) }\left( x;\omega ,a\right) =%
\mathit{L}_{n_{m}}^{i_{m}\alpha }\left( -i_{m}z\right) \frac{\Delta ^{\left(
n_{1},i_{1},,...,n_{m},i_{m}\right) }\left( x;\omega ,a\right) }{\Delta
^{\left( n_{1},i_{1},,...,n_{m-1},i_{m-1}\right) }\left( x;\omega ,a\right) }
\label{etatnonphys2}
\end{equation}%
is a rational function. Due to the presence of the exponential in $\phi
_{0,i_{m}}$, the behaviour at infinity of $\phi _{n_{m},i_{m}}^{\left(
n_{1},i_{1},...,n_{m-1},i_{m-1}\right) }$ is the same as for $\phi
_{0,i_{m}} $, namely

\begin{equation}
\left\{ 
\begin{array}{c}
\phi _{n_{m},+}^{\left( n_{1},i_{1},...,n_{m-1},i_{m-1}\right) }(x;\omega ,a)%
\underset{x\rightarrow +\infty }{\sim }\exp \left( z/2\right) x^{\alpha
+1/2}R^{\left( n_{1},i_{1},,...,n_{m},+\right) }\left( +\infty ;\omega
,a\right) \underset{x\rightarrow +\infty }{\rightarrow }\pm \infty \\ 
\phi _{n_{m},-}^{\left( n_{1},i_{1},...,n_{m-1},i_{m-1}\right) }(x;\omega ,a)%
\underset{x\rightarrow +\infty }{\sim }\exp \left( -z/2\right) x^{-\alpha
+1/2}R^{\left( n_{1},i_{1},,...,n_{m},-\right) }\left( +\infty ;\omega
,a\right) \underset{x\rightarrow +\infty }{\rightarrow }0^{\pm },%
\end{array}%
\right.  \label{CLinfini}
\end{equation}%
(the sign of the limit being the one of $R^{\left(
n_{1},i_{1},,...,n_{m},+\right) }\left( +\infty ;\omega ,a\right) $ or $%
R^{\left( n_{1},i_{1},,...,n_{m},-\right) }\left( +\infty ;\omega ,a\right) $
respectively).

The absence of node for $\phi _{n_{m},i_{m}}^{\left(
n_{1},i_{1},...,n_{m-1},i_{m-1}\right) }$ and the strict isospectrality
between $V^{\left( n_{1},i_{1},...,n_{m-1},i_{m-1}\right) }$ and $V^{\left(
n_{1},i_{1},...,n_{m},i_{m}\right) }$, are simultaneously ensured if and
only if we have

\begin{equation}
\left\{ 
\begin{array}{c}
\phi _{n_{m},+}^{\left( n_{1},i_{1},...,n_{m-1},i_{m-1}\right) }(x;\omega ,a)%
\underset{x\rightarrow 0^{+}}{\sim }x^{\alpha +1/2}R^{\left(
n_{1},i_{1},,...,n_{m},+\right) }\left( 0^{+};\omega ,a\right) \underset{%
x\rightarrow 0^{+}}{\rightarrow }0^{\pm } \\ 
\phi _{n_{m},-}^{\left( n_{1},i_{1},...,n_{m-1},i_{m-1}\right) }(x;\omega ,a)%
\underset{x\rightarrow 0^{+}}{\sim }x^{-\alpha +1/2}R^{\left(
n_{1},i_{1},,...,n_{m},-\right) }\left( 0^{+};\omega ,a\right) \underset{%
x\rightarrow 0^{+}}{\rightarrow }\pm \infty ,%
\end{array}%
\right.  \label{CL0}
\end{equation}%
(the sign of the limit being the one of $R^{\left(
n_{1},i_{1},,...,n_{m},+\right) }\left( 0^{+};\omega ,a\right) $ or $%
R^{\left( n_{1},i_{1},,...,n_{m},-\right) }\left( 0^{+};\omega ,a\right) $
respectively), with

\begin{equation*}
sign\left( R^{\left( n_{1},i_{1},,...,n_{m},i_{m}\right) }\left(
0^{+};\omega ,a\right) \right) =sign\left( R^{\left(
n_{1},i_{1},,...,n_{m},i_{m}\right) }\left( +\infty ;\omega ,a\right)
\right) .
\end{equation*}

\section{Two-step case}

Before to consider the general case, we first consider the two-step case. We
then have to check that the unphysical eigenstates $\phi
_{n_{2},i_{2}}^{\left( n_{1},i_{1}\right) }$ satisfy the appropriate
boundary conditions. Using Eq(\ref{L12states}) in Eq(\ref{etats niso}), we
have for $m=2$

\begin{equation*}
\phi _{n_{2},i_{2}}^{\left( n_{1},i_{1}\right) }(x;\omega ,a)=\frac{W\left(
\phi _{0,i_{1}}(x;\omega ,a)\mathit{L}_{n_{1}}^{i_{1}\alpha }\left(
-i_{1}z\right) ,\phi _{0,i_{2}}(x;\omega ,a)\mathit{L}_{n_{2}}^{i_{2}\alpha
}\left( -i_{2}z\right) \mid x\right) }{\phi _{0,i_{1}}(x;\omega ,a)\mathit{L}%
_{n_{1}}^{i_{1}\alpha }\left( -i_{1}z\right) },
\end{equation*}%
where $\alpha >-i_{j}n_{j},\ j=1,2$, or with Eq(\ref{factwronsk}) 
\begin{equation}
\phi _{n_{2},i_{2}}^{\left( n_{1},i_{1}\right) }(x;\omega ,a)=\phi
_{0,i_{1}}\left( x;\omega ,a\right) \frac{W\left( \mathit{L}%
_{n_{1}}^{i_{1}\alpha }\left( -i_{1}z\right) ,\left( x^{\alpha }\exp \left(
z/2\right) \right) ^{i_{2}-i_{1}}\mathit{L}_{n_{2}}^{i_{2}\alpha }\left(
-i_{2}z\right) \mid x\right) }{\mathit{L}_{n_{1}}^{i_{1}\alpha }\left(
-i_{1}z\right) }.  \label{wronsk2step}
\end{equation}

If $i_{1}=i_{2}=i$, then

\begin{equation}
\phi _{n_{2},i}^{\left( n_{1},i\right) }(x;\omega ,a)=\phi _{0,i}\left(
x;\omega ,a\right) \frac{W\left( \mathit{L}_{n_{1}}^{i\alpha }\left(
-iz\right) ,\mathit{L}_{n_{2}}^{i\alpha }\left( -iz\right) \mid x\right) }{%
\mathit{L}_{n_{1}}^{i\alpha }\left( -iz\right) }.
\end{equation}

Using \cite{muir}

\begin{equation}
W\left( y_{1},...,y_{m}\mid x\right) =\left( \frac{dz}{dx}\right)
^{m(m-1)/2}W\left( y_{1},...,y_{m}\mid z\right)  \label{changvarwronsk}
\end{equation}%
and \cite{magnus,szego}%
\begin{equation}
\left( \mathit{L}_{n}^{\alpha }\left( x\right) \right) ^{\prime }=-\mathit{L}%
_{n-1}^{\alpha +1}\left( x\right) ,  \label{derivlag}
\end{equation}%
this gives\bigskip

\begin{equation}
\phi _{n_{2},i}^{\left( n_{1},i\right) }(x;\omega ,a)=\frac{i\omega
x^{i\alpha +3/2}\exp \left( iz/2\right) }{\mathit{L}_{n_{1}}^{i\alpha
}\left( -iz\right) }\left\vert 
\begin{array}{cc}
\mathit{L}_{n_{1}}^{i\alpha }\left( -iz\right) & \mathit{L}_{n_{2}}^{i\alpha
}\left( -iz\right) \\ 
\mathit{L}_{n_{1}-1}^{i\alpha +1}\left( -iz\right) & \mathit{L}%
_{n_{2}-1}^{i\alpha +1}\left( -iz\right)%
\end{array}%
\right\vert .
\end{equation}

We then have (cf Eq(\ref{asymptLag}))

\begin{equation}
\left\{ 
\begin{array}{c}
\phi _{n_{2},i}^{\left( n_{1},i\right) }(x;\omega ,a)\underset{x\rightarrow
0^{+}}{\sim }\frac{i\omega x^{i\alpha +3/2}}{\binom{n_{1}+i\alpha }{n_{1}}}%
\left\vert 
\begin{array}{cc}
\binom{n_{1}+i\alpha }{n_{1}} & \binom{n_{2}+i\alpha }{n_{2}} \\ 
\binom{n_{1}+i\alpha }{n_{1}-1} & \binom{n_{2}+i\alpha }{n_{2}-1}%
\end{array}%
\right\vert =i\omega x^{i\alpha +3/2}\binom{n_{2}+i\alpha }{n_{2}-1}\frac{%
n_{2}-n_{1}}{n_{2}} \\ 
\phi _{n_{2},i}^{\left( n_{1},i\right) }(x;\omega ,a)\underset{x\rightarrow
+\infty }{\sim }i^{n_{2}}\left( n_{2}-n_{1}\right) \frac{\omega }{n_{2}!}%
x^{2n_{2}+i\alpha -1/2}\exp \left( iz/2\right) ,%
\end{array}%
\right.
\end{equation}%
that is,

\begin{equation}
\left\{ 
\begin{array}{c}
\phi _{n_{2},+}^{\left( n_{1},+\right) }(x;\omega ,a)\underset{x\rightarrow
0^{+}}{\sim }\omega x^{\left( \alpha +2\right) -1/2}\binom{n_{2}+\alpha }{%
n_{2}-1}\frac{n_{2}-n_{1}}{n_{2}}\underset{x\rightarrow 0^{+}}{\rightarrow }%
0^{\pm } \\ 
\phi _{n_{2},-}^{\left( n_{1},-\right) }(x;\omega ,a)\underset{x\rightarrow
+\infty }{\sim }\frac{\omega \left( n_{2}-n_{1}\right) }{n_{2}!}%
x^{2n_{2}+\alpha -1/2}\exp \left( z/2\right) \underset{x\rightarrow +\infty }%
{\rightarrow }\pm \infty ,%
\end{array}%
\right. ,\text{ if }a>1,  \label{asympt2step1}
\end{equation}%
where $\pm =sign(n_{2}-n_{1})$, and

\begin{equation}
\left\{ 
\begin{array}{c}
\phi _{n_{2},-}^{\left( n_{1},-\right) }(x;\omega ,a)\underset{x\rightarrow
0^{+}}{\sim }\left( -1\right) ^{n_{2}}\omega x^{-\left( \alpha -2\right)
-1/2}\binom{\alpha -n_{2}}{n_{2}-1}\frac{n_{2}-n_{1}}{n_{2}}\underset{%
x\rightarrow 0^{+}}{\rightarrow }\pm \infty \\ 
\phi _{n_{2},-}^{\left( n_{1},-\right) }(x;\omega ,a)\underset{x\rightarrow
+\infty }{\sim }\left( -1\right) ^{n_{2}}\frac{\omega \left(
n_{2}-n_{1}\right) }{n_{2}!}x^{2n_{2}-\alpha -1/2}\exp \left( -z/2\right) 
\underset{x\rightarrow +\infty }{\rightarrow }0^{\pm }%
\end{array}%
\right. ,\text{ if }a>\sup \left( n_{2},n_{1}\right) \text{ },
\label{asympt2step2}
\end{equation}%
where $\pm =\left( -1\right) ^{n_{2}}sign(n_{2}-n_{1})$.

If $i_{1}=-i_{2}=-i$, then using the last of the following identities \cite%
{askey}

\begin{equation}
\left\{ 
\begin{array}{c}
\left( z^{\alpha }\mathit{L}_{n}^{\alpha }\left( z\right) \right) ^{\prime
}=\left( n+\alpha \right) z^{\alpha -1}\mathit{L}_{n}^{\alpha -1}\left(
z\right) \\ 
\left( e^{-z}\mathit{L}_{n}^{\alpha }\left( z\right) \right) ^{\prime
}=-e^{-z}\mathit{L}_{n}^{\alpha +1}\left( z\right) \\ 
\left( z^{\alpha }e^{-z}\mathit{L}_{n}^{\alpha }\left( z\right) \right)
^{\prime }=\left( n+1\right) z^{\alpha -1}e^{-z}\mathit{L}_{n+1}^{\alpha
-1}\left( z\right) ,%
\end{array}%
\right.  \label{propLag}
\end{equation}%
we obtain

\begin{eqnarray}
\phi _{n_{2},i}^{\left( n_{1},-i\right) }(x;\omega ,a) &=&\omega x^{-i\alpha
+3/2}\exp \left( -iz/2\right) \left( \frac{2}{\omega }\right) ^{i\alpha }%
\frac{W\left( \mathit{L}_{n_{1}}^{-i\alpha }\left( iz\right) ,z^{i\alpha
}\exp \left( iz\right) \mathit{L}_{n_{2}}^{i\alpha }\left( -iz\right) \mid
z\right) }{\mathit{L}_{n_{1}}^{-i\alpha }\left( iz\right) } \\
&=&\frac{2x^{i\alpha -1/2}\exp \left( iz/2\right) }{\mathit{L}%
_{n_{1}}^{-i\alpha }\left( iz\right) }\left\vert 
\begin{array}{cc}
\mathit{L}_{n_{1}}^{-i\alpha }\left( iz\right) & z\mathit{L}%
_{n_{2}}^{i\alpha }\left( -iz\right) \\ 
-i\mathit{L}_{n_{1}-1}^{-i\alpha +1}\left( iz\right) & \left( n_{2}+1\right) 
\mathit{L}_{n_{2}}^{i\alpha -1}\left( -iz\right)%
\end{array}%
\right\vert .  \notag
\end{eqnarray}

We then have (cf Eq(\ref{asymptLag}))

\begin{equation}
\left\{ 
\begin{array}{c}
\phi _{n_{2},i}^{\left( n_{1},-i\right) }(x;\omega ,a)\underset{x\rightarrow
0^{+}}{\sim }2x^{i\alpha -1/2}i^{n_{2}+1}\frac{\left( \alpha +in_{2}\right)
...\alpha }{n_{2}!} \\ 
\phi _{n_{2},i}^{\left( n_{1},-i\right) }(x;\omega ,a)\underset{x\rightarrow
+\infty }{\sim }i^{n_{2}+1}\frac{1}{n_{2}!}\left( \frac{\omega }{2}\right)
^{n_{2}+1}x^{2n_{2}+i\alpha +3/2}\exp \left( iz/2\right) .%
\end{array}%
\right.
\end{equation}

Consequently ($a>1$)

\begin{equation}
\left\{ 
\begin{array}{c}
\phi _{n_{2},+}^{\left( n_{1},-\right) }(x;\omega ,a)\underset{x\rightarrow
0^{+}}{\sim }2x^{\alpha -1/2}\frac{\left( \alpha +n_{2}\right) ...\alpha }{%
n_{2}!}\underset{x\rightarrow 0^{+}}{\rightarrow }0^{+} \\ 
\phi _{n_{2},+}^{\left( n_{1},-\right) }(x;\omega ,a)\underset{x\rightarrow
+\infty }{\sim }\frac{1}{n_{2}!}\left( \frac{\omega }{2}\right)
^{n_{2}+1}x^{2n_{2}+a+1}\exp \left( z/2\right) \underset{x\rightarrow
+\infty }{\rightarrow }+\infty%
\end{array}%
\right.
\end{equation}%
and when $\alpha >n_{2}$

\begin{equation}
\left\{ 
\begin{array}{c}
\phi _{n_{2},-}^{\left( n_{1},+\right) }(x;\omega ,a)\underset{x\rightarrow
0^{+}}{\sim }\left( -1\right) ^{n_{2}+1}2\frac{\left( \alpha -n_{2}\right)
...\alpha }{n_{2}!}x^{-\alpha -1/2}\underset{x\rightarrow 0^{+}}{\rightarrow 
}\pm \infty \\ 
\phi _{n_{2},-}^{\left( n_{1},+\right) }(x;\omega ,a)\underset{x\rightarrow
+\infty }{\sim }\left( -1\right) ^{n_{2}+1}\frac{1}{n_{2}!}\left( \frac{%
\omega }{2}\right) ^{n_{2}+1}x^{2n_{2}-a+2}\exp \left( -z/2\right) \underset{%
x\rightarrow +\infty }{\rightarrow }0^{\pm },%
\end{array}%
\right.  \label{asympt2step3}
\end{equation}%
with $\pm =\left( -1\right) ^{n_{2}+1}$.

When $a>\sup (n_{2},n_{1})$, we have the correct asymptotic behaviour which
ensures the regularity of $V^{\left( n_{1},i_{1},n_{2},i_{2}\right)
}(x;\omega ,a)$ and $V^{\left( n_{1},i_{1}\right) }(x;\omega ,a)$ as their
strict isospectrality with $V(x;\omega ,a)$ for every choice of $\left(
n_{1},i_{1},n_{2},i_{2}\right) $ with $n_{1}\neq n_{2}$.

We then obtain a chain with two strictly isospectral regular successive
extensions of $V$

\begin{equation}
V\overset{A(v_{n_{1},i_{1}})}{\rightarrowtail }V^{\left( n_{1},i_{1}\right) }%
\overset{A(v_{n_{2},i_{2}}^{\left( n_{1},i_{1}\right) })}{\rightarrowtail }%
V^{\left( n_{1},i_{1},n_{2},i_{2}\right) }.
\end{equation}

\section{\protect\bigskip General m-step case}

Consider now the general case of a $m$-step DBT. We proceed by induction,
supposing that we have built a chain of $m-1$ regular strictly isospectral
potentials

\begin{equation}
V\overset{A(v_{n_{1},i_{1}})}{\rightarrowtail }V^{\left( n_{1},i_{1}\right) }%
\overset{A(v_{n_{2},i_{2}}^{\left( n_{1},i_{1}\right) })}{\rightarrowtail }%
...\overset{A(v_{n_{m-1},i_{m-1}}^{\left(
n_{1},i_{1},,...,n_{m-2},i_{m-2}\right) })}{\rightarrowtail }V^{\left(
n_{1},i_{1},...,n_{m-1},i_{m-1}\right) }.
\end{equation}%
and that we have $\alpha >-i_{j}n_{j},\ \forall j\leq m$. This implies in
particular that the $\phi _{n_{j},i_{j}}^{\left(
n_{1},i_{1},...,n_{j-1},i_{j-1}\right) }(x;\omega ,a),\ j\leq m-1$, which
are eigenfunctions of $\widehat{H}^{\left(
n_{1},i_{1},...,n_{m-1},i_{m-1}\right) }$ associated to the respective
negative eigenvalues $\mathcal{E}_{n_{j-1},i_{j-1}}(\omega ,a)<0$, are all
nodeless on $\left] 0,+\infty \right[ $ and that they keep a constant sign
on this interval that we can always take as positive.

We have to determine what is the constraint to implement for the unphysical
eigenfunction $\phi _{n_{m},i_{m}}^{\left(
n_{1},i_{1},...,n_{m-1},i_{m-1}\right) }$ in order that it satisfies the
good boundary conditions necessary to build a regular and strictly
isospectral extension $V^{\left( n_{1},i_{1},...,n_{m},i_{m}\right) }$ at
the next step.

From Eq(\ref{CLinfini}) we know that

\begin{equation}
\left\{ 
\begin{array}{c}
\phi _{n_{m-1},+}^{\left( n_{1},i_{1},...,n_{m-2},i_{m-2}\right) }(x;\omega
,a)\underset{x\rightarrow +\infty }{\rightarrow }+\infty \\ 
\phi _{n_{m-1},-}^{\left( n_{1},i_{1},...,n_{m-2},i_{m-2}\right) }(x;\omega
,a)\underset{x\rightarrow +\infty }{\rightarrow }0^{+},%
\end{array}%
\right.  \label{CLinfini2}
\end{equation}%
and consequently we have also (cf Eq(\ref{CL0}))

\begin{equation}
\left\{ 
\begin{array}{c}
\phi _{n_{m-1},+}^{\left( n_{1},i_{1},...,n_{m-2},i_{m-2}\right) }(x;\omega
,a)\underset{x\rightarrow 0^{+}}{\rightarrow }0^{+} \\ 
\phi _{n_{m-1},-}^{\left( n_{1},i_{1},...,n_{m-2},i_{m-2}\right) }(x;\omega
,a)\underset{x\rightarrow 0^{+}}{\rightarrow }+\infty .%
\end{array}%
\right.  \label{CL02}
\end{equation}

For $x$ sufficiently large or in the vinicity of $0$, we have $\mathcal{E}%
_{n_{m-1},\pm }(\omega ,a)<V^{\left( n_{1},i_{1},...,n_{m-2},i_{m-2}\right)
}(x;\omega ,a)$. The B\^{o}cher disconjugacy theorem then implies that in
the neighbourhoods of $0$ and of $+\infty $, the functions $\phi
_{n_{m-1},i_{m-1}}^{\left( n_{1},i_{1},...,n_{m-2},i_{m-2}\right) }$ are
strictly monotonous, that is strictly decreasing if $i_{m-1}=-$ and strictly
increasing if $i_{m-1}=+$.

We already know (cf Eq(\ref{CLinfini})) that

\begin{equation}
\phi _{n_{m},i_{m}}^{\left( n_{1},i_{1},...,n_{m-1},i_{m-1}\right) }(+\infty
;\omega ,a)=\left\{ 
\begin{array}{c}
\pm \infty ,\text{ if }i_{m}=+ \\ 
0^{\pm },\text{ if }i_{m}=-%
\end{array}%
\right.
\end{equation}

We prove now by induction the following result for the unphysical
eigenfunctions of the potentials obtained at the $\left( m-1\right) ^{th}$
step ($z=\omega x^{2}/2$)

\begin{equation}
\phi _{n_{m},i_{m}}^{\left( n_{1},i_{1},...,n_{m-1},i_{m-1}\right)
}(x;\omega ,a)\underset{x\rightarrow 0^{+}}{\sim }\widetilde{\phi }%
_{n_{m},i_{m}}^{\left( n_{1},i_{1},...,n_{m-1},i_{m-1}\right)
}(0)x^{i_{m}\left( \alpha +q\left( i_{1},...,i_{m}\right) \right)
-1/2}\left( 1+O(z)\right) ,  \label{compasympt}
\end{equation}%
where $\widetilde{\phi }_{n_{m},i_{m}}^{\left(
n_{1},i_{1},...,n_{m-1},i_{m-1}\right) }(0)$ has the same sign as $\phi
_{n_{m},i_{m}}^{\left( n_{1},i_{1},...,n_{m-1},i_{m-1}\right) }(+\infty
;\omega ,a)$ and where $q\left( i_{1},...,i_{m}\right) =q_{+}\left(
i_{1},...,i_{m}\right) -q_{-}\left( i_{1},...,i_{m}\right) $, $q_{\pm
}\left( i_{1},...,i_{m}\right) $ is the number of $i_{j}$ equal to $\pm $ in
the set $\left( i_{1},...,i_{m}\right) $. $q_{\pm }\left(
i_{1},...,i_{m}\right) $ can be viewed as the number of state of "$\pm $
charge" used in the chain of DBT at the $m^{th}$ step and the difference $%
q\left( i_{1},...,i_{m}\right) $ as the total charge associated to the chain.

We then have

\begin{equation}
\left\{ 
\begin{array}{c}
\phi _{n_{m},+}^{\left( n_{1},i_{1},...,n_{m-1},i_{m-1}\right) }(x;\omega ,a)%
\underset{x\rightarrow 0^{+}}{\rightarrow }0^{\pm },\text{ if }a>-q\left(
i_{1},...,i_{m-1},+\right) +1=-q\left( i_{1},...,i_{m-1}\right) \\ 
\phi _{n_{m},-}^{\left( n_{1},i_{1},...,n_{m-1},i_{m-1}\right) }(x;\omega ,a)%
\underset{x\rightarrow 0^{+}}{\rightarrow }\pm \infty ,\text{ if }a>-q\left(
i_{1},...,i_{m-1},-\right) =1-q\left( i_{1},...,i_{m-1}\right) ,%
\end{array}%
\right. \text{ }  \label{condnec}
\end{equation}%
which are necessary conditions in order that the potentials $V^{\left(
n_{1},i_{1},...,n_{m},+\right) }$ and $V^{\left(
n_{1},i_{1},...,n_{m},-\right) }$ respectively are regular and strictly
isospectral to $V^{\left( n_{1},i_{1},,...,n_{m-1},i_{m-1}\right) }$. Note
that these constraints are independent of the $n_{j}$ but depend only on the
type of the states used to build the chain.

Refering to the results obtained in the $2$-step case (see Eq(\ref%
{asympt2step1}), Eq(\ref{asympt2step2}) and Eq(\ref{asympt2step3})), we can
verify that this property is valid for $m=2$ since

\begin{equation}
\left\{ 
\begin{array}{c}
q\left( +,+\right) =-q\left( -,-\right) =2 \\ 
q\left( +,-\right) =q\left( -,+\right) =0.%
\end{array}%
\right.  \label{charge2step}
\end{equation}

Suppose that Eq(\ref{compasympt}) is verified till the $\left( m-1\right)
^{th}$ step, that is,

\begin{equation}
\phi _{n,\pm }^{\left( n_{1},i_{1},...,n_{m-2},i_{m-2}\right) }(x;\omega ,a)%
\underset{x\rightarrow 0^{+}}{\sim }x^{\pm \left( \alpha +q\left(
i_{1},...,i_{m-2}\right) \right) +1/2}\left( 1+O(z)\right) ,  \label{asympt2}
\end{equation}%
with $a>-q\left( i_{1},...,i_{m-2},+\right) =-1-q\left(
i_{1},...,i_{m-2}\right) $, if $i_{m-1}=+$, and $a>1-q\left(
i_{1},...,i_{m-2},-\right) =2-q\left( i_{1},...,i_{m-2}\right) $, if $%
i_{m-1}=-$.

\bigskip Since $\phi _{n_{m-1},i_{m-1}}^{\left(
n_{1},i_{1},...,n_{m-2},i_{m-2}\right) }$ and $\phi _{k,i}^{\left(
n_{1},i_{1},...,n_{m-2},i_{m-2}\right) }$ are both eigenfunctions of $%
\widehat{H}^{\left( n_{1},i_{1},...,n_{m-2},i_{m-2}\right) }$, using the
wronskian theorem \cite{messiah} in Eq(\ref{etats niso}) we can write the
unphysical eigenfunctions at the $\left( m-1\right) ^{th}$ step as

\begin{eqnarray}
\phi _{k,i}^{\left( n_{1},i_{1},...,n_{m-1},i_{m-1}\right) }(x;\omega ,a) &=&%
\frac{\left( \mathcal{E}_{n_{m-1},i_{m-1}}-\mathcal{E}_{k,i}\right)
\int_{x_{0}}^{x}d\xi \phi _{n_{m-1},i_{m-1}}^{\left(
n_{1},i_{1},...,n_{m-2},i_{m-2}\right) }(\xi ;\omega ,a)\phi _{k,i}^{\left(
n_{1},i_{1},...,n_{m-2},i_{m-2}\right) }(\xi ;\omega ,a)}{\phi
_{n_{m-1},i_{m-1}}^{\left( n_{1},i_{1},...,n_{m-2},i_{m-2}\right) }(x;\omega
,a)}  \label{wronskform} \\
&&+\frac{W\left( \phi _{n_{m-1},i_{m-1}}^{\left(
n_{1},i_{1},...,n_{m-2},i_{m-2}\right) },\phi _{k,i}^{\left(
n_{1},i_{1},...,n_{m-2},i_{m-2}\right) }\mid x_{0}\right) }{\phi
_{n_{m-1},i_{m-1}}^{\left( n_{1},i_{1},...,n_{m-1},i_{m-1}\right) }(x;\omega
,a)}.  \notag
\end{eqnarray}

We have to consider several cases corresponding to the four possible couples 
$\left( i_{m-1},i\right) $.

1) Consider first the case $i=i_{m-1}=+$ in which we take $x_{0}=0$. Using
Eq(\ref{CL02}), Eq(\ref{wronskform}) gives

\begin{equation}
\phi _{k,+}^{\left( n_{1},i_{1},...,n_{m-2},i_{m-2},n_{m-1},+\right)
}(x;\omega ,a)\sim \frac{\int_{0}^{x}d\xi \phi _{n_{m-1},+}^{\left(
n_{1},i_{1},...,n_{m-2},i_{m-2}\right) }(\xi ;\omega ,a)\phi _{k,+}^{\left(
n_{1},i_{1},...,n_{m-2},i_{m-2}\right) }(\xi ;\omega ,a)}{\phi
_{n_{m-1},+}^{\left( n_{1},i_{1},...,n_{m-2},i_{m-2}\right) }(x;\omega ,a)}.
\label{asympt++1}
\end{equation}

Since $\phi _{n_{m-1},+}^{\left( n_{1},i_{1},...,n_{m-2},i_{m-2}\right) }$
and $\phi _{k,+}^{\left( n_{1},i_{1},...,n_{m-2},i_{m-2}\right) }$ are both
strictly positive on $\left] 0,+\infty \right[ $, $\phi _{k,+}^{\left(
n_{1},i_{1},...,n_{m-2},i_{m-2},n_{m-1},+\right) }$ keeps a constant sign on
this interval and we can always take it as positive. Using Eq(\ref{asympt2}%
), Eq(\ref{asympt++1}) gives

\begin{equation}
\phi _{k,+}^{\left( n_{1},i_{1},...,n_{m-2},i_{m-2},n_{m-1},+\right)
}(x;\omega ,a)\underset{x\rightarrow 0^{+}}{\sim }x^{\left( \alpha +q\left(
i_{1},...,i_{m-2},+,+\right) \right) -1/2}\left( 1+O(z)\right) ,
\end{equation}

Consequently, in this case, Eq(\ref{compasympt}) is also verified at the $%
m^{th}$ step.

2) In the case $i=i_{m-1}=-$, if we take $x_{0}=+\infty $, we have similarly
with Eq(\ref{CLinfini2}) and Eq(\ref{wronskform})

\begin{equation}
\phi _{k,-}^{\left( n_{1},i_{1},...,n_{m-2},i_{m-2},n_{m-1},-\right)
}(x;\omega ,a)\sim \frac{\int_{x}^{+\infty }d\xi \phi _{n_{m-1},-}^{\left(
n_{1},i_{1},...,n_{m-2},i_{m-2}\right) }(\xi ;\omega ,a)\phi _{k,-}^{\left(
n_{1},i_{1},...,n_{m-2},i_{m-2}\right) }(\xi ;\omega ,a)}{\phi
_{n_{m-1},-}^{\left( n_{1},i_{1},...,n_{m-2},i_{m-2}\right) }(x;\omega ,a)}.
\label{asympt--1}
\end{equation}

As in the preceding case, since $\phi _{n_{m-1},-}^{\left(
n_{1},i_{1},...,n_{m-2},i_{m-2}\right) }$ and $\phi _{k,-}^{\left(
n_{1},i_{1},...,n_{m-2},i_{m-2}\right) }$ are both strictly positive on $%
\left] 0,+\infty \right[ $, $\phi _{k,-}^{\left(
n_{1},i_{1},...,n_{m-2},i_{m-2},n_{m-1},-\right) }$ keeps a constant sign on
this interval and we can always take it as positive. From Eq(\ref{CLinfini})
we deduce that for any $\varepsilon >0$, $\int_{\varepsilon }^{+\infty }d\xi
\phi _{n_{m-1},-}^{\left( n_{1},i_{1},...,n_{m-2},i_{m-2}\right) }(\xi
;\omega ,a)\phi _{k,-}^{\left( n_{1},i_{1},...,n_{m-2},i_{m-2}\right) }(\xi
;\omega ,a)$ is finite and positive. Since ($a>1-q\left(
i_{1},...,i_{m-2}\right) $)%
\begin{equation*}
\phi _{n_{m-1},-}^{\left( n_{1},i_{1},...,n_{m-2},i_{m-2}\right) }(x;\omega
,a)\phi _{k,-}^{\left( n_{1},i_{1},...,n_{m-2},i_{m-2}\right) }(x;\omega ,a)%
\underset{x\rightarrow 0^{+}}{\sim }x^{-2\left( \alpha +q\left(
i_{1},...,i_{m-2}\right) \right) +1}\left( 1+O(z)\right) ,
\end{equation*}%
Eq(\ref{asympt--1}) gives

\begin{equation}
\phi _{k,-}^{\left( n_{1},i_{1},...,n_{m-2},i_{m-2},n_{m-1},-\right)
}(x;\omega ,a)\underset{x\rightarrow 0^{+}}{\sim }x^{-\left( \alpha +q\left(
i_{1},...,i_{m-2},-,-\right) \right) -1/2}.  \notag
\end{equation}

Eq(\ref{compasympt}) is again verified at the $m^{th}$ step.

3) It remains to consider the cases $i=+=-i_{m-1}$ and $i=-=-i_{m-1}$. Since 
$\phi _{n,+}^{\left( n_{1},i_{1},...,n_{m-2},i_{m-2}\right) }$ and $\phi
_{n,-}^{\left( n_{1},i_{1},...,n_{m-2},i_{m-2}\right) }$ have not the same
behavior nor at $0$ nor at $+\infty $, Eq(\ref{wronskform}) is not adapted
and we proceed to a direct evaluation of the wronskian $W\left( \phi
_{n_{m-1},-}^{\left( n_{1},i_{1},...,n_{m-2},i_{m-2}\right) },\phi
_{k,+}^{\left( n_{1},i_{1},...,n_{m-2},i_{m-2}\right) }\mid x\right) $.

Since $\phi _{n_{m-1},-}^{\left( n_{1},i_{1},...,n_{m-2},i_{m-2}\right) }$
and $\phi _{k,+}^{\left( n_{1},i_{1},...,n_{m-2},i_{m-2}\right) }$ are
strictly positive and respectively decreasing and increasing both at
infinity and in the vinicity of the origin, $W\left( \phi
_{n_{m-1},-}^{\left( n_{1},i_{1},...,n_{m-2},i_{m-2}\right) },\phi
_{k,+}^{\left( n_{1},i_{1},...,n_{m-2},i_{m-2}\right) }\mid x\right) $ has
the same positive sign at these two limits. Due to the disconjugacy of

\begin{equation}
\phi ^{\prime \prime }(x;\omega ,a)+\left( \mathcal{E}_{k,+}(\omega
,a)-V^{\left( n_{1},i_{1},...,n_{m-1},i_{m-1}\right) }(x;\omega ,a)\right)
\phi (x;\omega ,a)=0,
\end{equation}%
we then deduce that $\phi _{k,+}^{\left(
n_{1},i_{1},...,n_{m-2},i_{m-2},n_{m-1},-\right) }(x;\omega ,a)$ keeps a
constant sign (taken as positive) on all the interval $\left] 0,+\infty %
\right[ $.

Using (see Eq(\ref{asympt2}))

\begin{equation}
\left\{ 
\begin{array}{c}
\phi _{k,+}^{\left( n_{1},i_{1},...,n_{m-2},i_{m-2}\right) }(x;\omega ,a)%
\underset{x\rightarrow 0^{+}}{\sim }x^{\left( \alpha +q\left(
i_{1},...,i_{m-2}\right) \right) +1/2}\left( \widetilde{\phi }_{k,+}^{\left(
n_{1},i_{1},...,n_{m-2},i_{m-2}\right) }(0)+O(z)\right) \\ 
\phi _{n_{m-1},-}^{\left( n_{1},i_{1},...,n_{m-2},i_{m-2}\right) }(x;\omega
,a)\underset{x\rightarrow 0^{+}}{\sim }x^{-\left( \alpha +q\left(
i_{1},...,i_{m-2}\right) \right) +1/2}\left( \widetilde{\phi }%
_{n_{m-1},-}^{\left( n_{1},i_{1},...,n_{m-2},i_{m-2}\right) }(0)+O(z)\right)
,%
\end{array}%
\right.  \label{asympt1}
\end{equation}%
we obtain

\begin{eqnarray}
&&W\left( \phi _{n_{m-1},-}^{\left( n_{1},i_{1},...,n_{m-2},i_{m-2}\right)
},\phi _{k,+}^{\left( n_{1},i_{1},...,n_{m-2},i_{m-2}\right) }\mid x\right) 
\notag \\
&&\underset{x\rightarrow 0^{+}}{\sim }2\left( \alpha +q\left(
i_{1},...,i_{m-2}\right) \right) \widetilde{\phi }_{k,+}^{\left(
n_{1},i_{1},...,n_{m-2},i_{m-2}\right) }(0)\widetilde{\phi }%
_{n_{m},-}^{\left( n_{1},i_{1},...,n_{m-2},i_{m-2}\right) }(0)+O(z).
\end{eqnarray}

Consequently, we deduce

\begin{equation}
\phi _{k,+}^{\left( n_{1},i_{1},...,n_{m-2},i_{m-2},n_{m-1},-\right)
}(x;\omega ,a)\underset{x\rightarrow 0^{+}}{\sim }x^{\left( \alpha +q\left(
i_{1},...,i_{m-2},-,+\right) \right) -1/2}  \label{asympt-+}
\end{equation}%
and

\begin{eqnarray}
&&\phi _{k,-}^{\left( n_{1},i_{1},...,n_{m-2},i_{m-2},n_{m-1},+\right)
}(x;\omega ,a)=\frac{W\left( \phi _{n_{m-1},+}^{\left(
n_{1},i_{1},...,n_{m-2},i_{m-2}\right) },\phi _{k,-}^{\left(
n_{1},i_{1},...,n_{m-2},i_{m-2}\right) }\mid x\right) }{\phi
_{n_{m},+}^{\left( n_{1},i_{1},...,n_{m-2},i_{m-2}\right) }(x;\omega ,a)}
\label{asympt+-4} \\
&&\underset{x\rightarrow 0^{+}}{\sim }-2\left( \alpha +q\left(
i_{1},...,i_{m-2}\right) \right) \widetilde{\phi }_{k,-}^{\left(
n_{1},i_{1},...,n_{m-2},i_{m-2}\right) }(0)x^{-\left( \alpha +q\left(
i_{1},...,i_{m-2}\right) \right) -1/2}\left( 1+O(z)\right)  \notag \\
&&\underset{x\rightarrow 0^{+}}{\sim }x^{-\left( \alpha +q\left(
i_{1},...,i_{m-2},+,-\right) \right) -1/2}.  \notag
\end{eqnarray}

The result Eq(\ref{compasympt}) is then still valid at the $m^{th}$ step.

We still have to verify that this condition is compatible with the
regularity of the preceding potentials in the chain. For that, note that at
the $m^{th}$ step the explicit expression of the resulting potential and its
associated eigenfunctions are a priori independent of the ordering chosen
for the sequence . This means that we can consider several chains leading to
the same extended potential. Consider the set $\left(
n_{1},i_{1},...,n_{m-2},i_{m-2},n_{m-1},i_{m-1}\right) $ in which we have
regrouped first all the $n_{j}$ associated to a positive $i_{j}$, that is,
we have re-arranged the chain as $\left( n_{1}^{\prime },i_{1}^{\prime
},...,n_{q_{+}}^{\prime },i_{q_{+}}^{\prime },n_{q_{+}+1}^{\prime
},i_{q_{+}+1}^{\prime },...,n_{m-1}^{\prime },i_{m-1}^{\prime }\right)
=\left( n_{1}^{\prime },+,...,n_{q_{+}}^{\prime },+,n_{q_{+}+1}^{\prime
},-,...,n_{m-1}^{\prime },-\right) $, where $\left( n_{1}^{\prime
},...,n_{m-1}^{\prime }\right) $ is a permutation of $\left(
n_{1},...,n_{m-1}\right) $ and where $q_{+}=q_{+}\left(
i_{1},...,i_{m-1}\right) $.

Suppose first that $q_{+}=m-1$, that is, $i_{j}^{\prime }=+,\ \forall j\in
\left\{ 1,...,m-1\right\} $. $\forall j\in \left\{ 1,...,m-1\right\} $ we
have $1-q_{+}\left( i_{1}^{\prime },...,i_{j}^{\prime }\right) \leq 0$ and
since $a>1$, the $\phi _{n_{j+1}^{\prime },i_{j+1}^{\prime }}^{\left(
n_{1}^{\prime },i_{1}^{\prime },...,n_{j}^{\prime },i_{j}^{\prime }\right) }$
are all nodeless implying the regularity of all the potentials $V^{\left(
n_{1}^{\prime },i_{1}^{\prime },...,n_{j}^{\prime },i_{j}^{\prime }\right) }$
in the chain.

Suppose now that $q_{+}<m-1$, that is, $i_{m-1}^{\prime }=-$. The condition $%
a>1-q\left( i_{1}^{\prime },...,i_{m-1}^{\prime }\right) =2-q\left(
i_{1}^{\prime },...,i_{m-2}^{\prime }\right) >1-q\left( i_{1}^{\prime
},...,i_{m-2}^{\prime }\right) $ implies the absence of node for $\phi
_{n_{m-1}^{\prime },-}^{\left( n_{1}^{\prime },i_{1}^{\prime
},...,n_{m-2}^{\prime },i_{m-2}^{\prime }\right) }$ and consequently the
regularity of $V^{\left( n_{1}^{\prime },i_{1}^{\prime },...,n_{m-1}^{\prime
},i_{m-1}^{\prime }\right) }$.

We can iterate this reasoning to go backward along the chain and we see
clearly that at each step, the opposite of the "total charge" decreases,
ensuring the regularity of the associated potential. In this procedure, once
we attain the level $q_{+}$, we have eliminated all the negative $%
i_{j}^{\prime }$ and retrieve then the first case.

Consequently, if we satisfy the condition $a>1-q\left(
i_{1},...,i_{m-1}\right) $ at the step $m$, we are ensured of the existence
of a chain of regular extensions which leads to a regular $V^{\left(
n_{1},i_{1},,...,n_{m},i_{m}\right) }$.

\section{Eigenstates of the extended potentials}

We can generalize the calculations made in the two-step case to determine
explicit expressions for the eigenstates of the successive extensions. For
that we adopt the ordering mentioned above and consider the spectrum of the
potential $V^{\left(
n_{1},+,...,n_{q_{+}},+,n_{q_{+}+1},-,....,n_{m},-\right) }$ where $%
q_{+}=q_{+}\left( i_{1},...,i_{m}\right) $. Using Eq(\ref{factwronsk}) and
Eq(\ref{changvarwronsk}), we have

\begin{eqnarray}
&&W\left( \phi _{n_{1},+},...,\phi _{n_{q_{+}},+},...,\phi
_{n_{q_{+}+1},-},...,\phi _{n_{m},-},\psi _{k}\mid x\right) \\
&\sim &x^{\left( m+1\right) \left( m/2+a\right) }e^{-\left( m+1\right)
z/2}W\left( e^{z}L_{n_{1}}^{\alpha }\left( -z\right)
,...,e^{z}L_{n_{q_{+}}}^{\alpha }\left( -z\right) ,z^{-\alpha
}L_{n_{q_{+}+1}}^{-\alpha }\left( z\right) ,...,z^{-\alpha
}L_{n_{m}}^{-\alpha }(z),L_{k}^{\alpha }\left( z\right) \mid x\right) , 
\notag
\end{eqnarray}%
that is, with Eq(\ref{derivlag}) and Eq(\ref{propLag})

\begin{eqnarray}
&&W\left( \phi _{n_{1},+},...,\phi _{n_{q_{+}},+},...,\phi
_{n_{q_{+}+1},-},...,\phi _{n_{m},-},\psi _{k}\mid x\right) \\
&\sim &x^{\left( m+1\right) \left( m/2+a\right) }e^{\left( q_{+}-\left(
m+1\right) /2\right) z}z^{-(m-q_{+})(\alpha +m)}\det \left( \Psi
_{k}^{\left( n_{1},+,...,n_{q_{+}},+,n_{q_{+}+1},-,....,n_{m},-\right)
}\right) ,  \notag
\end{eqnarray}%
where ($i,j\in \left\{ 1,...,m+1\right\} $)

\begin{equation}
\left[ \Psi _{k}^{\left(
n_{1},+,...,n_{q_{+}},+,n_{q_{+}+1},-,....,n_{m},-\right) }\right]
_{i,j}=\left\{ 
\begin{array}{c}
L_{n_{j}}^{\alpha +i-1}\left( -z\right) ,\text{ if }j\leq q_{+} \\ 
\left( n_{j}-\alpha -i+2\right) _{i-1}z^{m-i+1}L_{n_{j}}^{-\alpha -i+1}(z),%
\text{ if }q_{+}+1\leq j\leq m \\ 
\left( -1\right) ^{i-1}L_{k-i+1}^{\alpha +i-1}\left( z\right) ,\text{ if }%
j=m+1.%
\end{array}%
\right. \ 
\end{equation}

In the same manner

\begin{eqnarray}
&&W\left( \phi _{n_{1},+},...,\phi _{n_{q_{+}},+},...,\phi
_{n_{q_{+}+1},-},...,\phi _{n_{m},-}\mid x\right) \\
&\sim &x^{m\left( \left( m-1\right) /2+a\right) }e^{mz/2}W\left(
L_{n_{1}}^{\alpha }\left( -z\right) ,...,L_{n_{q_{+}}}^{\alpha }\left(
-z\right) ,z^{-\alpha }e^{-z}L_{n_{q_{+}+1}}^{-\alpha }\left( z\right)
,...,z^{-\alpha }e^{-z}L_{n_{m}}^{-\alpha }(z)\mid x\right) ,  \notag
\end{eqnarray}%
that is,

\begin{eqnarray}
&&W\left( \phi _{n_{1},+},...,\phi _{n_{q_{+}},+},...,\phi
_{n_{q_{+}+1},-},...,\phi _{n_{m},-}\mid x\right) \\
&\sim &x^{m\left( \left( m-1\right) /2+a\right) }z^{(m-q_{+})\left( -\alpha
-m+1)\right) }e^{-\left( m/2-q_{+}\right) z}\det \left( \Phi ^{\left(
n_{1},+,...,n_{q_{+}},+,n_{q_{+}+1},-,....,n_{m},-\right) }\right) ,  \notag
\end{eqnarray}%
where ($i,j\in \left\{ 1,...,m\right\} $)

\begin{equation}
\left[ \Phi ^{\left(
n_{1},+,...,n_{q_{+}},+,n_{q_{+}+1},-,....,n_{m},-\right) }\right]
_{i,j}=\left\{ 
\begin{array}{c}
L_{n_{j}-i+1}^{\alpha +i-1}\left( -z\right) ,\text{ if }j\leq q_{+} \\ 
\left( n_{j}+1\right) _{i-1}z^{m-i+1}L_{n_{j}+i-1}^{-\alpha -i+1}(z),\text{
if }q_{+}+1\leq j\leq m.%
\end{array}%
\right. \ 
\end{equation}%
\bigskip

Eq(\ref{etats niso}) gives then

\begin{equation}
\psi _{k}^{\left( n_{1},+,...,n_{q_{+}},+,n_{q_{+}+1},-,....,n_{m},-\right)
}(x;\omega ,a)\sim \psi _{0}(x;\omega ,a+q\left( i_{1},...,i_{m}\right) )%
\frac{\det \left( \Psi _{k}^{\left(
n_{1},+,...,n_{q_{+}},+,n_{q_{+}+1},-,....,n_{m},-\right) }\right) }{\det
\left( \Phi ^{\left(
n_{1},+,...,n_{q_{+}},+,n_{q_{+}+1},-,....,n_{m},-\right) }\right) }.
\end{equation}

The $\psi _{k}^{\left( n_{1},i_{1},...,n_{m},i_{m}\right) }$ are then
obtained as the product of the modified "gauge factor" (ie fundamental state
of the initial potential) $\psi _{0}(x;\omega ,a+q\left(
i_{1},...,i_{m}\right) )$, multiplied by the ratio of two polynomials, the
polynomial denominator $\det \left( \Phi ^{\left(
n_{1},+,...,n_{q_{+}},+,n_{q_{+}+1},-,....,n_{m},-\right) }\right) $ being
common to all the eigenstates of the considered extension. The orthogonality
condition on the $\psi _{k}^{\left( n_{1},i_{1},...,n_{m},i_{m}\right) }$
ensure that the $\det \left( \Psi _{k}^{\left(
n_{1},+,...,n_{q_{+}},+,n_{q_{+}+1},-,....,n_{m},-\right) }\right) $
constitute an orthogonal family with respect to the weight

\begin{equation*}
W^{\left( n_{1},i_{1},...,n_{m},i_{m}\right) }(z)=\frac{z^{\alpha +q\left(
i_{1},...,i_{m}\right) }e^{-z}}{\left( \det \left( \Phi ^{\left(
n_{1},+,...,n_{q_{+}},+,n_{q_{+}+1},-,....,n_{m},-\right) }\right) \right)
^{2}}.
\end{equation*}

\section{Shape invariance of the extended potentials}

Consider a chain of regular strictly isospectral extensions

\begin{equation}
V\overset{A(v_{n_{1},i_{1}})}{\rightarrowtail }V^{\left( n_{1},i_{1}\right) }%
\overset{A(v_{n_{2},i_{2}}^{\left( n_{1},i_{1}\right) })}{\rightarrowtail }%
...\overset{A(v_{n_{m},i_{m}}^{\left(
n_{1},i_{1},,...,n_{m-1},i_{m-1}\right) })}{\rightarrowtail }V^{\left(
n_{1},i_{1},,...,n_{m},i_{m}\right) },
\end{equation}

The superpartner of the potential $V^{\left(
n_{1},i_{1},...,n_{m},i_{m}\right) }(x;\omega ,a)$ is given by \cite%
{cooper,Dutt}

\begin{equation}
\widetilde{V}^{\left( n_{1},i_{1},...,n_{m},i_{m}\right) }(x;\omega
,a)=V^{\left( n_{1},i_{1},...,n_{m},i_{m}\right) }(x;\omega ,a)+2\left(
w_{0}^{\left( n_{1},i_{1},...,n_{m},i_{m}\right) }(x;\omega ,a)\right)
^{\prime },  \label{SUSYpart}
\end{equation}%
$w_{0}^{\left( n_{1},i_{1},...,n_{m},i_{m}\right) }(x;\omega ,a)$ being the
RS function associated to the ground level of $V^{\left(
n_{1},i_{1},...,n_{m},i_{m}\right) }$ ($E_{0}\left( \omega \right) =0$).

Since (see Eq(\ref{transfoback2}))

\begin{equation}
w_{0}^{\left( n_{1},i_{1},...,n_{m},i_{m}\right) }(x;\omega
,a)=-v_{n_{m},i_{m}}^{\left( n_{1},i_{1},...,n_{m-1},i_{m-1}\right)
}(x;\omega ,a)-\frac{\mathcal{E}_{n_{m},i_{m}}(\omega ,a)}{%
v_{n_{m},i_{m}}^{\left( n_{1},i_{1},...,n_{m-1},i_{m-1}\right) }(x;\omega
,a)-w_{0}^{\left( n_{1},i_{1},...,n_{m-1},i_{m-1}\right) }(x;\omega ,a)},
\label{transfoback3}
\end{equation}%
we have with Eq(\ref{potnstepisot})

\begin{eqnarray}
\widetilde{V}^{\left( n_{1},i_{1},...,n_{m},i_{m}\right) }(x;\omega ,a)
&=&V^{\left( n_{1},i_{1},...,n_{m-1},i_{m-1}\right) }(x;\omega ,a)
\label{SUSYpart1} \\
&&-2\left( \frac{\mathcal{E}_{n_{m},i_{m}}(\omega ,a)}{v_{n_{m},i_{m}}^{%
\left( n_{1},i_{1},...,n_{m-1},i_{m-1}\right) }(x;\omega ,a)-w_{0}^{\left(
n_{1},i_{1},...,n_{m-1},i_{m-1}\right) }(x;\omega ,a)}\right) ^{\prime }.
\end{eqnarray}

We proceed by induction. Suppose that the potential $V^{\left(
n_{1},i_{1},...,n_{m-1},i_{m-1}\right) }$ has the same shape invariance
properties than $V$

\begin{eqnarray}
\widetilde{V}^{\left( n_{1},i_{1},...,n_{m-1},i_{m-1}\right) }(x;\omega ,a)
&=&V^{\left( n_{1},i_{1},...,n_{m-1},i_{m-1}\right) }(x;\omega ,a)+2\left(
w_{0}^{\left( n_{1},i_{1},...,n_{m-1},i_{m-1}\right) }(x;\omega ,a)\right)
^{\prime } \\
&=&V^{\left( n_{1},i_{1},...,n_{m-1},i_{m-1}\right) }(x;\omega ,a+1)+2\omega
.  \notag
\end{eqnarray}

As shown in \cite{grandati3}, this is effectively the case for $m=2$. It
results

\begin{equation}
\widetilde{V}^{\left( n_{1},i_{1},...,n_{m},i_{m}\right) }(x;\omega
,a)=V^{\left( n_{1},i_{1},...,n_{m},i_{m}\right) }(x;\omega ,a+1)+2\omega
-2\left( \Delta ^{\left( n_{1},i_{1},...,n_{m},i_{m}\right) }(x;\omega
,a)\right) ^{\prime },  \label{susym}
\end{equation}%
where

\begin{eqnarray}
\Delta ^{\left( n_{1},i_{1},...,n_{m},i_{m}\right) }(x;\omega ,a) &=&\frac{%
\mathcal{E}_{n_{m},i_{m}}(\omega ,a)}{v_{n_{m},i_{m}}^{\left(
n_{1},i_{1},...,n_{m-1},i_{m-1}\right) }(x;\omega ,a)-w_{0}^{\left(
n_{1},i_{1},...,n_{m-1},i_{m-1}\right) }(x;\omega ,a)}  \label{deltam} \\
&&+w_{0}^{\left( n_{1},i_{1},...,n_{m-1},i_{m-1}\right) }(x;\omega
,a)+v_{n_{m},i_{m}}^{\left( n_{1},i_{1},...,n_{m-1},i_{m-1}\right)
}(x;\omega ,a+1).  \notag
\end{eqnarray}

Suppose also that

\begin{equation}
\Delta ^{\left( n_{1},i_{1},...,n_{m-1},i_{m-1}\right) }(x;\omega ,a)=0,
\end{equation}%
which again is verified for $m=2$ \cite{grandati3}. We can then write

\begin{eqnarray}
v_{n_{m-1},i_{m-1}}^{\left( n_{1},i_{1},...,n_{m-2},i_{m-2}\right)
}(x;\omega ,a+1) &=&-w_{0}^{\left( n_{1},i_{1},...,n_{m-2},i_{m-2}\right)
}(x;\omega ,a)  \label{propm-1} \\
&&-\frac{\mathcal{E}_{n_{m-1},i_{m-1}}\left( \omega ,a\right) }{%
v_{n_{m-1},i_{m-1}}^{\left( n_{1},i_{1},...,n_{m-2},i_{m-2}\right)
}(x;\omega ,a)-w_{0}^{\left( n_{1},i_{1},...,n_{m-2},i_{m-2}\right)
}(x;\omega ,a)}.  \notag
\end{eqnarray}

Moreover we have

\begin{equation}
\left\{ 
\begin{array}{c}
w_{0}^{\left( n_{1},i_{1},...,n_{m-1},i_{m-1}\right) }(x;\omega
,a)=-v_{n_{m-1},i_{m-1}}^{\left( n_{1},i_{1},...,n_{m-2},i_{m-2}\right)
}(x;\omega ,a)-\frac{\mathcal{E}_{n_{m-1},i_{m-1}}\left( \omega ,a\right) }{%
v_{n_{m-1},i_{m-1}}^{\left( n_{1},i_{1},...,n_{m-2},i_{m-2}\right)
}(x;\omega ,a)-w_{0}^{\left( n_{1},i_{1},...,n_{m-2},i_{m-2}\right)
}(x;\omega ,a)} \\ 
v_{n_{m},i_{m}}^{\left( n_{1},i_{1},...,n_{m-1},i_{m-1}\right) }(x;\omega
,a)=-v_{n_{m-1},i_{m-1}}^{\left( n_{1},i_{1},...,n_{m-2},i_{m-2}\right)
}(x;\omega ,a)+\frac{\mathcal{E}_{n_{m},i_{m}}\left( \omega ,a\right) -%
\mathcal{E}_{n_{m-1},i_{m-1}}\left( \omega ,a\right) }{v_{n_{m-1},i_{m-1}}^{%
\left( n_{1},i_{1},...,n_{m-2},i_{m-2}\right) }(x;\omega
,a)-v_{n_{m},i_{m}}^{\left( n_{1},i_{1},...,n_{m-2},i_{m-2}\right)
}(x;\omega ,a)}%
\end{array}%
\right.  \label{backm-1}
\end{equation}%
and (see Eq(\ref{Energiesneg}))

\begin{equation}
\mathcal{E}_{n_{m},i_{m}}\left( \omega ,a+1\right) -\mathcal{E}%
_{n_{m-1},i_{m-1}}\left( \omega ,a+1\right) =\mathcal{E}_{n_{m},i_{m}}\left(
\omega ,a\right) -\mathcal{E}_{n_{m-1},i_{m-1}}\left( \omega ,a\right) .
\label{deltaenerg}
\end{equation}

Inserting Eq(\ref{backm-1}), Eq(\ref{propm-1}) and Eq(\ref{deltaenerg}) into
Eq(\ref{deltam}), we obtain after a little elementary algebra

\begin{equation}
\Delta ^{\left( n_{1},i_{1},...,n_{m},i_{m}\right) }(x;\omega ,a)=0.
\end{equation}

By induction, this property is verified for every $m$. Then Eq(\ref{susym})
becomes

\begin{equation}
\widetilde{V}^{\left( n_{1},i_{1},...,n_{m},i_{m}\right) }(x;\omega ,a)=%
\widetilde{V}^{\left( n_{1},i_{1},...,n_{m},i_{m}\right) }(x;\omega
,a+1)+2\omega ,
\end{equation}%
that is, the potential has also the same shape invariance properties than
the isotonic potential. The translational shape invariance of the isotonic
potential is then hereditary in such chain of extensions.

\section{Conclusion}

We have shown that the "generalized SUSY QM partnership" that we have
previously elaborated in a one-step scheme can be extended in a multi-step
formulation. We have proven the necessary conditions to obtain chains of
regular extensions of the isotonic potential of arbitrary length and have
given explicit expressions for their eigenstates. We also established
explicitely the hereditary character of the shape invariance properties of
the isotonic potential which are common to all the potentials in a given
chain.

The case of the generic potentials of the second category of primary TSIP 
\cite{grandati}, namely P\"{o}schl-Teller or Scarf potentials, can be
considered in the same way. This work is in progress and a forthcoming paper
is in preparation. For the first category exceptional TSIP, due to the
strict isospectrality constraint, the only interesting case in which we can
envisage such chain of extensions is the ERKC potential \cite{grandati4}.
This is the object of further investigations.

\section{Acknowledgments}

I would like to thank A.\ B\'{e}rard, R.\ Milson and C.\ Quesne for
stimulating exchanges and very interesting suggestions.

\end{document}